\documentclass[12pt,a4paper]{article}

\usepackage[scale={.75,.8}]{geometry} 
\usepackage{pstricks}
\usepackage[dvips]{graphicx}
\usepackage[latin2]{inputenc}
\usepackage{epic}
\usepackage{latexsym}
\usepackage{epsf}
\usepackage{epsfig}
\usepackage{amssymb,amsmath}
\usepackage{cite} 

\newcommand{\eref}[1]{eq.\ (\ref{e.#1})} 
\newcommand{\erefn}[1]{ (\ref{e.#1})}
 
\newcommand{\aref}[1]{\ref{a.#1}}
\newcommand{\sref}[1]{Section \ref{s.#1}}
\newcommand{\cref}[1]{Chapter \ref{c.#1}}

\def\nn{\nonumber \\}  
\newcommand{\nl}{& \nonumber \\ &}

\def\ds{\displaystyle}

\def\beq{\begin{equation}} 
\def\eeq{\end{equation}} 
\def\bea{\begin{eqnarray}}  
\def\eea{\end{eqnarray}}  
\newcommand{\bal}{\begin{align}}
\newcommand{\eal}{\end{align}}   
\def\ba{\begin{array}}  
\def\ea{\end{array}}   
\def\bi{\begin{itemize}}  
\def\ei{\end{itemize}}  
\def\ben{\begin{enumerate}}  
\def\een{\end{enumerate}}  
\def\beq{\begin{equation}}  
\def\eeq{\end{equation}}  
\def\bc{\begin{center}}
\def\ec{\end{center}} 
 \def\bt{\begin{table}}
\def\et{\end{table}}  
 \def\btb{\begin{tabular}}
\def\etb{\end{tabular}}  
\newcommand{\refeq}[1]{\mbox{(\ref{#1})}}

\def\cl{{\mathcal L}}  
   
\def\cm{{\mathcal M}}

\def\mass2{mass${}^2$}

\def\ads{{\mathrm A \mathrm d \mathrm S}}

\def\ra{\rangle}
\def\la{\langle}  

\def\pa{\partial}

\def\rt{\sqrt{2}} 
\def\simlt{\stackrel{<}{{}_\sim}}
\def\simgt{\stackrel{>}{{}_\sim}}

\newcommand{\llra}{\longleftrightarrow}

\newcommand{\oh}{\frac{1}{2}}

\begin{document}

\pagestyle{empty}
\begin{flushright}
CERN-PH-TH/2006-088\\ 
UG-FT-205/06 \\
CAFPE-75/06 \\

{\bf \today}
\end{flushright}
\vspace*{5mm}
\begin{center}

\renewcommand{\thefootnote}{\fnsymbol{footnote}}

{\large {\bf Tools for Deconstructing Gauge Theories in AdS${}_5$
}} \\ 
\vspace*{1cm}
{\bf Jorge~de~Blas}$^{\rm a)}$\footnote{Email:deblasm@ugr.es},
{\bf Adam~Falkowski}$^{\rm b),c)}$\footnote{Email:adam.falkowski@cern.ch},
{\bf Manuel~Perez-Victoria}$^{\rm b)}$\footnote{Email:mpv@mail.cern.ch}
and {\bf Stefan~Pokorski}$^{\rm c)}$\footnote{Email:pokorski@fuw.edu.pl}
\vspace{0.5cm}

a) CAFPE and Departamento de Fisica Teorica y del Cosmos, \\ Universidad de Granada, E-18071, Spain \\
b) CERN Theory Division, CH-1211 Geneva 23, Switzerland\\
c) Institute of Theoretical Physics, Warsaw University, \\ Ho\.za 69, 00-681 Warsaw, Poland

\vspace*{1.7cm}
{\bf Abstract}
\end{center}
\vspace*{5mm}
\noindent
{ 
We employ analytical methods to study deconstruction of 5D gauge theories in the $\ads_5$ background. 
We demonstrate that using the so-called  q-Bessel functions allows a quantitative analysis of the deconstructed setup.
Our study clarifies the relation of deconstruction with 5D warped theories.
}
\vspace*{1.0cm}
\date{\today}


\vspace*{0.2cm}
 
\vfill\eject
\newpage

\setcounter{page}{1}
\pagestyle{plain}

\renewcommand{\thefootnote}{\arabic{footnote}}
\setcounter{footnote}{0}

\section{Introduction}

The framework introduced by  Randall and Sundrum in ref.\ \cite{rasu} has 
been  playing an important  role in high-energy physics in the last
years. 
The Randall--Sundrum setup involves 5D spacetime with the line element  
\beq
\label{e.wbi}
ds^2 = a^2(x_5) dx_\mu dx^\mu  - dx_5^2 \, , 
\eeq
and the fifth dimension truncated  at $x_5 =0$  by the ultraviolet 
(UV) brane and at $x_5 = L$ by the infrared (IR) brane.
The scale (warp) factor multiplying the 4D Minkowski background metric 
varies along  the fifth dimension, 
which generates a hierarchy of scales  of order $a(0)/a(L)$ between the UV and 
the IR branes.
For example, for the $\ads_5$ spacetime the warp factor varies 
exponentially and a huge hierarchy can easily be generated.  

The original motivation was to explain in this way the hierarchy between 
the Planck scale and the electroweak scale.
However it has become clear that the scope of application is much wider.
In particular, it is interesting to consider situations where the 
gravitational degrees of freedom can be  decoupled.
In this case one deals with  5D theories of gauge and matter fields in 
a fixed background of the form of \eref{wbi}.     
This approach has turned out to be fruitful for constructing realistic
models of the Higgs sector \cite{higgs},  Higgsless electroweak 
breaking \cite{higgsless} and supersymmetry breaking \cite{susyads}.   
Furthermore, AdS/CFT~\cite{adscft} applied to the Randall--Sundrum 
background \cite{adscftrs} suggests that these models are dual 
descriptions of purely four-dimensional strongly coupled physics.
There is also a connection between field-theoretical 5D
gauge theories in warped backgrounds and the low-energy physics of
pions and vector resonances \cite{adsqcd}, known as AdS/QCD.

Gauge theories in $D>4$ dimensions possess an intrinsic cutoff scale.
The gauge coupling $g_D$ has dimension $[{\rm length}]^{D/2-2}$;
therefore at high energies scattering amplitudes grow as
$E^{D/2-2}$, leading to strong coupling in the UV.  
Perturbative computations have to be cut off below the strong coupling 
scale.
In some phenomenological applications strong coupling occurs, in fact, 
not far from the TeV scale.
Thus it is often desirable to consider a UV completion of higher 
dimensional theories so as to understand possible  cutoff effects.  
Typically, this UV completion is assumed to be some sort of string
theory. 

Deconstruction \cite{dec} is another option to model cutoff effects in  
higher dimensional gauge theories.
It is a four-dimensional framework that typically involves a product 
gauge group $G^N$ and a set of bifundamental Higgs fields (the links).
With an appropriate choice of  representations and vev's of the links, such a 
setup, at low energies, reproduces the spectrum and interactions of a 
higher-dimensional gauge theory with the gauge group $G$.
The matching holds up to a certain deviation scale $\Lambda_D$, related 
to the magnitude of the link vev's.   
This deviation scale is identified with the cutoff of the 
higher-dimensional theory.
Above $\Lambda_D$, deconstruction provides a UV completion in terms of a 
purely four-dimensional, weakly coupled and, possibly, renormalizable 
gauge theory dynamics.   

Deconstruction of 5D gauge theories in the $\ads_5$ Randall--Sundrum 
background was considered in refs. \cite{abkoma,faki,rashwe,kash,caergl}.
The warped fifth dimension was  represented by a chain of 
bifundamental links $\Phi_j$ with the vev varying as $\la \Phi_j \ra
\sim q^j$.   
This setup turned out to be useful for clarifying several issues 
concerning the evolution of gauge couplings in $\ads_5$ 
\cite{faki,rashwe,kash}.  
However, these studies suffer from certain limitations.
While computations  in $\ads_5$ can be performed (both at the
tree level and at the loop level) using familiar Bessel functions, 
no analytical methods have been available so far to handle  computations in
the deconstructed model. 
In particular, the spectrum and interactions of the massive gauge
bosons have been determined only numerically.  
Some analytical results have been obtained, but only for $q \ll 1$, in
which case deconstruction does not have an obvious 5D interpretation. 
For this reason the relation between deconstruction and the $\ads_5$
theory was somewhat obscure. It was not even clear if deconstruction
could really reproduce the $\ads_5$ physics in its entire perturbativity
range. Furthermore, because of these technical problems, the
deconstructed $\ads_5$ setup has not been really useful for
phenomenological applications.  

In this paper we fill this gap.
We present analytical methods to handle deconstructed $\ads_5$
models. The tools are provided by the mathematical theory of
q-difference equations and their solutions.
One of its branches deals with the so-called q-Bessel functions, which 
generalize the ordinary Bessel functions. We show that the q-Bessel
functions are appropriate to describe the spectrum and interactions of
the deconstructed $\ads_5$ models. With
these methods at hand, the calculabilities of the $\ads_5$ gauge
theory and its deconstructed version stand on an equal footing. 
   
The technical results we obtain help to clarify the relation  between
the 5D warped gauge theories and deconstruction. 
It was argued in the previous works \cite{faki,rashwe} that
deconstruction realizes a position-dependent cutoff.   
With the new methods at our disposal we are able to make this notion
more precise and specify the parameter range in which deconstruction
adequately approximates the 5D theory. 
The position-dependent cutoff is realized in the following sense.
The IR brane  scattering amplitudes  in the 5D theory are reproduced up 
to the scale $\Lambda_{\mathrm IR} = \Lambda a(L)$.
On the other hand,  the UV brane t-channel amplitudes are reproduced up to
$\Lambda_{\mathrm UV} = \Lambda a(0)$, larger by the factor $a(0)/a(L)$.
The non-trivial thing about the latter result is that the spectra of 
massive excitations in the two theories deviate at a much lower scale, 
of order $\Lambda_{\mathrm IR}$.
Matching of the UV brane t-channel amplitudes in the two theories holds 
in spite of the fact that the 
number and the couplings of the exchanged massive gauge bosons are 
different. 
Our results also imply that, up to the scale $\Lambda_{\mathrm UV}$, the holographic interpretation
of the deconstructed $\ads_5$ models is similar to that established in \cite{adscftrs} for the 5D models.

The paper is  organized as follows.
In Section 2 we review the necessary technical material concerning gauge 
theories in $\ads_5$.
In Section 3 we study the deconstructed model and derive analytical 
formulas for the spectrum and propagators of the gauge bosons.
The detailed discussion of these results and their consequences is presented in Section 4.
Section 5 contains a summary and points at future applications of our 
results. 
Deconstruction of 5D gauge theories with most general boundary conditions is discussed in \aref{dbc}.  
Finally,  \aref{qb} contains a detailed and self-contained review of the theory of q-Bessel functions.

\section{Review of gauge theory in ${\rm AdS}_5$}

In this section we summarize basic properties of 5D gauge theories in the Randall--Sundrum-type background.
The fifth dimension is an interval (equivalently, the orbifold $S^1/Z_2$) parametrized by  $x_5 \in [0,L]$. 
Two boundary branes  are located at $x_5 = 0$ (the UV brane) and at $x_5 = L$ (the IR brane).  
The gravitational  background is that of the $\ads_5$ geometry:
\beq
\label{e.wb}
ds^2 = a^2(x_5) \eta_{\mu\nu} dx_\mu dx_\nu  - dx_5^2 
\qquad \qquad
 a(x_5) =  e^{- k  x_5} \, ,
\eeq
where $k$ is the curvature scale and $\mu,\nu, \dots = 0 \dots 3$ label the 4D coordinates. 
The presence of the warp factor
 generates a hierarchy of scales  of order ${1 / a_L}$ between the UV and the IR brane, where 
$ a_L  \equiv a(L) = e^{- k L }$. 
In the original Randall--Sundrum model $a_L \sim 10^{-15}$, so as to explain  the hierarchy between the electroweak and the Planck scale.
In our analysis we allow for arbitrary $a_L \ll 1$.  
 
In the following we will focus on  the dynamics of gauge theories on
such a background, while fluctuations of the 5D metric  will be
ignored.
The practical reason is that deconstruction of gravity encounters
certain technical problems \cite{argesc,rascth}, which we prefer to
avoid here. Formally speaking, gravity can always be decoupled by taking the limit $M_5 \to \infty$, where $M_5$ is the 5D Planck scale. 
More precisely, gravity couples strongly at the IR brane at the scale 
$\Lambda_G(L) \sim  a_L M_{\mathrm Planck}  \sim  a_L M_5^{3/2}/k^{1/2}$,
while at the UV brane it becomes strongly coupled only at
$\Lambda_G(0) \sim  M_{\mathrm Planck}$.  
We assume we always deal with energies far below the respective
gravity strong-coupling scales, and concentrate exclusively on the
dynamics of the gauge sector.

We consider a 5D gauge field $A_M = (A_\mu,A_5)$ propagating in the background of \eref{wb}. 
The quadratic part of the action  takes the form
\beq
\label{e.wga}
S_5 = \int d^4 x \int_0^{L} dx_5 {1 \over g_5^2} \left( 
- {1 \over 4} F_{\mu\nu} F_{\mu\nu} +  {a^2(x_5) \over 2} (\pa_5 A_\mu - \pa_\mu A_5)^2  \right) \, . 
\eeq
We choose the boundary conditions as 
\beq
\pa_5 A_\mu(0) = \pa_5 A_\mu(L)  = 0 \qquad \qquad
A_5(0) = A_5(L) = 0 \, .
\eeq
The Neumann boundary conditions for $A_\mu$ allow the existence of a 4D zero mode, so that 4D gauge symmetry survives below the compactification scale.  
Alternatively, one can impose Dirichlet  boundary conditions for $A_\mu$ (and Neumann for $A_5$), in which case  gauge symmetry  is entirely broken at low energies.
For completeness we review this case and its deconstructed version in \aref{dbc}. 

\subsection{Spectrum and propagators}

There are several ways of approaching higher dimensional theories.
Most often, higher dimensional fields are traded for an infinite number of 4D Kaluza--Klein (KK) modes.
Another approach relies on 5D propagators, which effectively sum up
the contribution of the whole KK tower. 
Finally, it is sometimes useful to employ the so-called holographic approach, which consists in constructing  an effective action  for the boundary values of the bulk fields.
Below we review the necessary technical background for each approach.
Of course, these are merely different methods of organizing computations,  and physical conclusions must be the same, regardless of which approach is used.   

\subsubsection{Kaluza--Klein approach}

The gauge field is decomposed into KK modes as follows 
\beq
A_\mu(x,x_5) = \sum_{n =0}^\infty f_n(x_5) A_\mu^n(x) \qquad , \qquad 
A_5(x,x_5) = \sum_{n = 1}^\infty {\pa_5 f_n(x_5) \over m_n} A_5^n(x) \, .
\eeq
The modes $A_5^n$ are eaten by the massive gauge fields but it is not necessary to fix the gauge at this stage. 
The KK profiles $f_n$ should be  chosen such  that the quadratic action \erefn{wga} is diagonal in the KK basis:
\beq
S_5 = \int d^4 x  \sum_n \left(  
- {1 \over 4} F_{\mu\nu}^n F_{\mu\nu}^n  + {1 \over 2} (m_n A_\mu^n -  \pa_\mu A_5^n)^2  
\right) \, .
\eeq
This is obtained when $f_n$ are solutions  of the  equation of motion
\beq
\label{e.wfeom}
\pa_5^2 f_n - 2 k \pa_5 f_n + m_n^2 a^{-2}(x_5) f_n = 0 \, ,
\eeq
satisfy the Neumann boundary conditions
\beq
\label{e.wfbc}
\pa_5 f_n(0) = \pa_5 f_n (L) = 0 \, ,
\eeq
and are normalized as $g_5^{-2} \int_0^{L}  f_n^2 =  1$. 
The solution is given by \cite{poads5}
\bea 
\label{e.kkp}
\ds f_0  & = &  {g_5 \over \sqrt{L}} \, ,
\nn \ds
f_n & =& A_n a^{-1}(x_5) 
\left [ Y_0 \left ({m_n \over k }\right) J_1 \left({m_n \over k a(x_5)}\right) 
- J_0\left({m_n \over k }\right)Y_1\left({m_n \over k a(x_5)}\right) \right ] \, ,
 \qquad n >0 \, ,
\eea
with the normalization constant\footnote{%
We thank K.y.Oda for discussions on this and subsequent formulas in this subsection.  
}  
\beq 
A_n =  {\pi m_n g_5 \over \sqrt{2 k}} 
\left ( \left ({J_0({m_n \over k }) \over  J_0({m_n \over k a_L})}\right )^2 -1  \right )^{-1/2}
= {\pi m_n g_5 \over \sqrt{2 k}} 
\left ( \left ({Y_0({m_n \over k }) \over  Y_0({m_n \over k a_L})}\right )^2 -1  \right )^{-1/2} \, .
\eeq  

Thus, the 5D gauge theory is rewritten in terms of a tower of 4D vector fields $A_\mu^n$ with masses $m_n$ 
($A_5^n$'s are eaten).  
The zero mode $A_\mu^0$ corresponds to $m_0 = 0$. 
Its profile $f_0$ is a constant, thus the massless mode couples to all matter with equal strength given by $g_0 = g_5/\sqrt{L}$. 
For the massive modes, the mass spectrum  is  given by solutions of the equation  
\beq
\label{e.gbqc}
Y_0 \left ({m_n \over k }\right ) J_0\left({m_n \over k a_L}\right)  - J_0\left({m_n \over k }\right)Y_0\left({m_n \over k a_L}\right) = 0  \, ,
\eeq
which can be well approximated by 
\beq
\label{e.ckks}
m_n \approx \pi k a_L (n - 1/4) \, .
\eeq
The spectrum is approximately linearly spaced with the  mass gap  $M_{\mathrm KK} \sim k a_L$.  
The massive modes couple non-universally to matter, the coupling  depending on  both the KK number and the position in the extra dimension. 
In particular, the couplings to the UV and IR branes are determined by the boundary values of the KK profiles
\beq
f_n(0) = {\sqrt{2 k} g_5 \over 
\sqrt{\left ({Y_0({m_n \over k }) \over  Y_0({m_n \over k a_L})}\right
  )^2 -  1 }} \, ,
\qquad \qquad
f_n(L) =  
{\sqrt{2 k} g_5 \left ({Y_0({m_n \over k }) \over  Y_0({m_n \over k a_L})} \right )
\over 
\sqrt{\left ({Y_0({m_n \over k }) \over  Y_0({m_n \over k a_L})}\right
  )^2 -1 }}\, .
\eeq 
Approximating the Bessel functions we find 
$|f_n(L)|^2 \approx 2 k g_5^2$, so that all the KK modes couple to the IR brane with approximately equal strength. 
On the other hand 
$|f_n(0)|^2 \approx 2 k a_L g_5^2 (\pi k/2 m_n) \log^{-2} (m_n/k )$  for $m_n \ll  k$ 
and  $|f_n(0)|^2 \approx 2 k a_L g_5^2$ for $m_n \gg  k$. 
This shows that KK modes are localized toward the IR brane and couple very weakly to the UV brane. 
For this reason physics on the UV brane may remain perturbative at energies much larger than the KK mass gap.  

\subsubsection{Position-space approach} 

An alternative approach to quantum computations in higher dimensions
relies on 5D propagators. Since 4D Poincar\'e invariance is preserved,
it is convenient to work in a mixed representation of 4D momentum space
and position space in the extra dimension which, for brevity, we call
the position space representation: 
\beq
P_{\mu \nu} (p^2,x_5,y_5) =
\int d^4 x e^{i p x}  \la T A_\mu(x,x_5) A_\nu(0,y_5) \ra \, .
\eeq
At tree level the propagator is an inverse of the kinetic operator (Fourier transformed to 4D momentum space). 
We choose the gauge-fixing term 
$\cl_{\mathrm gf} = -{1 \over 2 \xi g_5^2}(\pa_\mu A_\mu - \xi \pa_5 (a^2(x_5) A_5))^2$.  
Then the propagator satisfies the equation 
\beq
{1 \over g_5^2} \left ( [-p^2 - \pa_5(a^2(x_5) \pa_5)]\eta_{\mu\nu}  + (1 - {1/\xi}) p_\mu p_\nu \right )
P_{\nu\rho}(p^2,x_5,y_5) =  i \eta_{\mu\rho} \delta(x_5 - y_5)
\label{e.propeq}
\eeq
and the Neumann boundary conditions 
$\pa_{x_5} P_{\mu \nu}(p^2,x_5,y_5)|_{x_5 = 0} =\pa_{x_5} P_{\mu \nu}(p^2,x_5,y_5)|_{x_5 = L} = 0$.   
The solution is given by \cite{rasc}
\beq
P_{\mu\nu}(p^2,x_5,y_5) = 
\left (\eta_{\mu\nu}  - {p_\mu p_\nu \over p^2} \right ) P (p^2,x_5,y_5)  + {p_\mu p_\nu \over p^2}P (p^2/\xi,x_5,y_5)
\eeq 
where
\beq
\label{e.pp}
P(p^2,x_5,y_5) = {i \pi g_5^2 \over 2 k}
{[Y_0({p \over k })J_1({p \over k a(x_5)}) -  J_0({p \over k })Y_1({p \over k a(x_5)})]
[Y_0({p \over k a_L})J_1({p \over k a(y_5)}) -  J_0({p \over k a_L})Y_1({p \over k a(y_5)})]
\over
a(x_5) a(y_5) [Y_0({p \over k })J_0({p \over k a_L}) -  J_0({p \over k })Y_0({p \over k a_L})]}
\eeq
for $x_5 \leq y_5$ (the propagator is of course symmetric under the exchange $x_5 \leftrightarrow y_5$). 

We will focus on  brane-to-brane propagators defined as 
\beq
P_{\mathrm UV}(p^2) \equiv P(p^2,0,0)\, ,  \qquad 
P_{\mathrm IR}(p^2) \equiv P(p^2,L,L)\, ,  \qquad 
P_{\mathrm BB}(p^2) \equiv P(p^2,0,L)\, . 
\eeq
They read
\bea
\label{e.bbp}
\ds P_{\mathrm UV}(p^2) &=& \ds  {i g_5^2 \over  p}
{Y_0({p \over k a_L})J_1({p \over k }) -  J_0({p \over k a_L})Y_1({p \over k })
\over
Y_0({p \over k })J_0({p \over k a_L}) -  J_0({p \over k })Y_0({p \over k a_L})} \, ,
\nn
\ds P_{\mathrm IR}(p^2) &=& \ds {i g_5^2 \over a_L p} 
{Y_0({p \over k })J_1({p \over k a_L}) -  J_0({p \over k })Y_1({p \over k a_L})
\over
Y_0({p \over k })J_0({p \over k a_L}) -  J_0({p \over k })Y_0({p \over k a_L})} \, ,
\nn
\ds P_{\mathrm BB}(p^2) &=& \ds   {2 i k g_5^2 \over \pi p^2} 
{1 \over
Y_0({p \over k })J_0({p \over k a_L}) -  J_0({p \over k })Y_0({p \over k a_L})}  \, .
\eea
Consider the Euclidean momenta $p^2 = - p_E^2$. 
For  $p_E \ll  k a_L$ all propagators approximate those of a 4D massless gauge boson with the gauge coupling $g_0 = g_5/\sqrt{L}$:  
\beq
\label{e.bbp4}
P_{\mathrm UV}(-p_E^2) \approx P_{\mathrm IR}(-p_E^2)  \approx P_{\mathrm BB}(-p_E^2)  \approx 
 {i g_0^2 \over p_E^2} \, .
\eeq
Above the KK mass gap the momentum dependence of the propagators crucially depends on the position in the extra dimension.
For $a_L k  \ll p_E \ll k$ one finds 
\bea
\label{e.bbpw}
P_{\mathrm UV}(-p_E^2) & \approx & {i g_5^2\over p_E} 
{K_1({p_E \over k })
\over 
 K_0({p_E \over k})} 
 \approx 
 {1 \over \log(2 k e^{-\gamma}/p_E)} {i g_5^2 k \over p_E^2} \, , 
\nn
P_{\mathrm IR}(-p_E^2) & \approx & {i  g_5^2 \over a_L p_E} \, ,
\nn
P_{\mathrm BB}(-p_E^2) & \approx & 
\sqrt{\pi k \over 2 a_L} {g_5^2 \over \log(2 k e^{-\gamma} /p_E)} {i \over p_E^{3/2}} 
 e^{-{p_E \over a_L k}} \, .
\eea
The propagator on the IR brane shows a $1/E$ fall-off, which is the same behaviour as for 5D flat spacetime. 
Propagation between the branes is exponentially suppressed, which is also a characteristic feature of higher dimensional theories above the compactification scale.  
But on the UV brane the fifth dimension is screened: 
the propagator exhibits  the usual 4D $1/E^2$ behaviour, up to a
``classical running'' encoded in the logarithmic form factor.  
Such behaviour persists all the way up to the curvature scale.
At even higher energies, for $p_E \gg  k$:
\beq
\label{e.bbpf}
P_{\mathrm UV}(-p_E^2)  \approx  {i g_5^2\over p_E} 
{K_1({p_E \over k })
\over 
 K_0({p_E \over k})}  \approx {i g_5^2 \over  p_E}   
\, , \qquad
P_{\mathrm IR}(-p_E^2)  \approx  {i g_5^2\over a_L p_E} 
\, , \qquad
P_{\mathrm BB}(-p_E^2) \approx  {i g_5^2\over a_L^{1/2} p_E}  e^{-{p_E \over
    a_L k}} 
\, .
\eeq
At very high energies, larger than the curvature scale, the propagators have an energy dependence  analogous to that of the propagators in the 5D flat spacetime. The UV and IR brane propagators differ only by  the inverse warp factor multiplying the latter.

The connection between the KK and the 5D position space approach is given by the spectral formula
\beq
P(p^2,x_5,y_5) = - i \, \sum_n \frac{f_n(x_5) f_n(y_5)} {p^2 - m_n^2} \, . \label{e.resonantsum} 
\eeq
This shows that the position propagator describes collective propagation  of all KK modes between $x_5$ and $y_5$.  
The KK masses and wave functions are given, respectively, by the poles and residua of the propagator.
The position propagator is thus a very convenient object, which  encodes
information about the whole KK spectrum. Moreover, locality in the 5th
dimension is explicit in this formalism. For this reason, 
it is particularly useful for the computation of scattering amplitudes of
fields localized on 4D branes. 

\subsubsection{Holographic approach} 

In the holographic approach we single out the UV boundary value of the 5D field, treating it as a distinct variable from the bulk or the IR brane values.  
The latter degrees of freedom are integrated out, leaving a non-local effective action
$S_{\mathrm{eff}}$ for the UV boundary value. 
At the classical level, integrating out amounts to evaluating the 5D
action on a field configuration that satisfies the bulk equations of
motion and the IR boundary conditions, with the UV brane value
$\bar{A}_\mu(p)$ left as the ``low energy variable''. 
In the gauge $\xi = 1$, such a configuration can be written as 
\beq
A_\mu(p,x_5)=   {P(p^2,x_5,0) \over P(p^2,0,0)}  \bar{A}_\mu(p) \, , 
\eeq
where the propagator $P(p^2,x_5,y_5)$ was defined in \eref{pp}. 
Inserting this into the 5D action we obtain  the effective
action for $\bar{A}_\mu(p)$:
\beq
\label{e.h}
S_{\mathrm{eff}} = - \frac{1}{2} \int {d^4 p \over (2\pi)^4}
 \bar{A}_\mu(p) \Pi_{\mu \nu} \bar{A}_\nu(p) \, ,  \qquad 
 \Pi_{\mu \nu} = -{i \over P_{\mathrm UV}(p^2)} \eta_{\mu\nu} \, .
\eeq
We see that, at tree level, the self-energy of  $\bar{A}_\mu(p)$ is an inverse of the  UV brane-to-brane propagator in~\refeq{e.bbp}. 

In principle, this procedure can be applied to arbitrary geometries,
whenever the boundary value is for some reason a convenient low energy
variable.\footnote{This is the case, for example, in models with the
  SM gauge fields living in the bulk and the SM matter localized on
  the UV brane \cite{baporast}.}   
However for the background of \eref{wb} the AdS/CFT correspondence~\cite{adscft} applied to the Randall--Sundrum geometry \cite{adscftrs} gives it a special meaning.
The gauge theory we study here is dual to some 4D strongly coupled CFT, with a 4D gauge field weakly coupled to a conserved current $J_\mu$ of the CFT.   
Hence $\Pi_{\mu\nu}$ represents the connected correlator of two conserved currents, $J_\mu$ and $J_\nu$, in the dual 4D theory. 
This allows us to understand the peculiar features of the UV propagator described in the previous subsection.  
In particular, the form of $P_{\mathrm UV}$ in the regime 
$k a_L \ll p_E \ll k$ (see Eq.~\refeq{e.bbpw}) is fixed by  conformal invariance of the 4D holographic theory.

Cutting off $\ads_5$  with the UV brane is interpreted as  an explicit breaking of
conformal invariance in the 4D dual  by  a UV cutoff of order $k$. 
On the other hand, the presence of the IR brane is interpreted as a  spontaneous breaking of conformal
invariance in the 4D theory. 
This results in a mass gap of order $k a_L$ and in  a discrete spectrum of resonances
that are  identified (up to a mixing with $\bar{A}_\mu$) with the KK modes in the 5D picture. 
According to AdS/CFT, the  tree-level approximation of the 5D theory corresponds to a large-$N$ limit in the dual theory. 
In fact, it is well known that, in the large-$N$ limit, the exact two-point correlator function can be written as a sum of infinitely narrow resonances, in agreement with~\refeq{e.resonantsum}.

\subsection{Scales} 

We close this section with a discussion of the energy range where
perturbative gauge theories can be applied. As
we are interested in the parametric dependence only, we do not display
any numerical factors explicitly.

So far we have encountered two scales:  
the curvature scale $k$  and the KK scale $M_{\mathrm KK} \sim  a_L k$, that mark a qualitative change in the behaviour of the propagators.  
In 5D gauge theories the gauge coupling has dimension $[{\rm length}]^{1/2}$,
 therefore another scale $1/g_5^2$ appears. 
This quantity is related to the strong coupling scale of the theory. 
A simple way to see this is by coupling the gauge field to  4D matter sectors localized on the branes, say, to massless fermions. 
At tree level, $t$-channel amplitudes for two-by-two scatterings of the brane fields 
depend on  the scattering energy as $\cm_{UV,IR} \sim  E^2 P_{UV,IR}(-E^2)$.
Below the KK scale  the theory is effectively four-dimensional, thus the amplitudes do not grow with energy.  
One finds $\cm_{\mathrm UV,\mathrm IR} \sim g_5^2/L = g_0^2$ and, of course, we assume that the zero mode gauge  coupling $g_0$ is perturbative.
Once we cross the KK scale the IR brane-to-brane propagator switches to a 5D behaviour. 
The IR amplitudes then  grow as   $\cm_{\mathrm IR} \sim a_L^{-1}E g_5^2$ and  violate the unitarity bound at 
$E \sim \Lambda_S(L) = a_L/g_5^{2}$.
Above $\Lambda_S(L)$ the IR brane fields are strongly coupled and the perturbative description of the IR physics is no longer valid. 
On the other hand, for $M_{\mathrm KK} < E < k$ the UV amplitudes evolve only logarithmically, 
$\cm_{\mathrm UV} \sim g_5^2 k/\log(k/E)$ (we assume that the amplitude remains perturbative in this energy regime).
The linear growth starts only above the curvature scale:
 for $E > k$  we find  $\cm_{\mathrm UV} \sim E g_5^2$ leading to the strong coupling at 
$E \sim \Lambda_S(0) = 1/g_5^{2}$. 

We can generalize the preceding arguments by inserting 4D test matter fields at an arbitrary position in the fifth dimension. 
The obvious outcome is that the strong coupling scale depends on the position as
\beq
\Lambda_S(x_5) = {a(x_5) \over g_5^2} \, .
\eeq 
The strong coupling scale sets a limit on the validity range of the gauge theory. 
We are forced to cut off perturbative computations below $\Lambda_S$ and assume 
that some UV completion (or a non-perturbative formulation) properly describes the physics above $\Lambda_S$. 
Since the maximum validity range depends on the position, it is natural to consider a position-dependent cutoff, $\Lambda(x_5) = a(x_5) \Lambda$, with $\Lambda \simlt 1/g_5^2$.    
Deconstruction provides a framework to introduce such a cutoff in a gauge-invariant way. 

We have argued that UV brane physics remains perturbative at energies
much higher than the strong coupling scale on the IR brane. 
This is possible because of the locality of higher dimensional gauge theories.   
At virtualities $E$ bigger than the KK mass gap, the gauge bosons
propagate only a distance of order $ E^{-1}$ into the bulk, and simply
do not reach the strongly coupled region close to the IR brane, as can be
seen explicitly in the approximate Euclidean brane-to-brane propagator
of~\refeq{e.bbpf}. 
In fact, for $E > M_{\mathrm KK}$ the Euclidean UV brane propagators would
remain essentially unchanged if the IR brane were removed, that is for
$L \to \infty$. 

There is however a caveat.
For time-like momenta, the UV brane propagators have poles at the
position of the KK masses. At the quantum level, these poles will
correspond to resonances with finite widths. Measuring these widths, a
UV observer can determine whether there is strong coupling somewhere
else, in particular in the IR brane. This requires at least an energy
resolution of the order of the KK mass splittings. We can understand
this in an alternative way by looking at the propagator between the two
branes, which is no longer 
exponentially suppressed for time-like momenta, even in the regime $p
\gg k a_L$.  Instead, 
it has an oscillatory behaviour, so that in principle a UV brane observer
is able to probe the strong coupling of the IR brane physics.  
In practice, however, because the UV scattered particles have non-zero
energy spread~$\Gamma$, one must average over this range of energies,
and the interference of the different modes leads to an exponential 
suppression. This effect can be taken into
account by evaluating the propagators at $\sqrt{p^2} = 2E + i \Gamma$,
which leads to a suppression factor $\sim \exp(-\Gamma/a_L
k)$ \cite{ra}.
 Hence, we see again that the strongly coupled IR brane physics
remains screened from the UV brane observers, as long as energy
resolution is worse than the 
spacing between KK modes $\sim a_L k$.

\section{Deconstruction}
\label{s.de}
We move to the deconstruction of  warped gauge theories
\cite{abkoma,faki,rashwe}.  
Our goal for this section is to perform an analytical computation of
the spectrum and  propagators in a model approximating the gauge
theory in the Randall--Sundrum background. 
A detailed physical  discussion of the results  is postponed to \sref{di}.  
For definiteness, we restrict in the following to deconstructing 5D theories with $U(1)$ gauge group.  
For $U(n)$ or $SU(n)$ the analysis is very similar and  poses no additional technical problems.  

We  consider a $U(1)^{N+1}$ gauge theory with  $N$ complex scalars (the {\it links}) $\Phi_j$ with charges  $(1_{j-1}, - 1_j)$.
The action is given by
\beq
\label{renaction}
S_4  = \int d^4 x  \left( 
- {1 \over 4 g^2} \sum_{j = 0}^{N} F_{\mu\nu}^j F_{\mu\nu}^j 
 +  \sum_{j = 1}^{N}   |\pa_\mu \Phi_j + i  (A_\mu^{j-1} - A_\mu^j) \Phi_j |^2   + V(|\Phi^j|^2)
 \right) \, .
\eeq
The gauge coupling of each gauge group is set equal to $g$. 
Moreover, at tree level we forbid kinetic operators $F_{\mu \nu}^j F_{\mu\nu}^k$ with $j \neq k$ or higher-dimensional operators coupling different links. 
These arbitrary choices are meant to reproduce  certain features of the 5D theory such as 5D coordinate invariance and locality.
Furthermore, we assume that the scalar potential is such that all the links acquire vev's, 
$\la \Phi_j \ra  = v_j/\sqrt{2}$.
One real component of each $\Phi_j$ gets mass of order $v_j$ while the other stays massless 
(it is the Goldstone boson eaten by the gauge field that becomes massive).
We isolate the massless components and ignore the massive ones by going
to a non-linear parametrization, 
$\Phi_j \to  {v_j \over \rt} e^{i G_j/v_j}$, in which the action depends only on derivatives of $G_j$.
The action for $A_\mu^j$ and $G_j$ becomes 
\beq
\label{e.dqa}
S_4 \to  \int d^4 x  \left( 
- {1 \over 4 g^2} \sum_{j = 0}^{N} F_{\mu\nu}^j F_{\mu\nu}^j 
 +  \sum_{j = 1}^{N}  {1 \over 2} \left[ v_j (A_\mu^{j} - A_\mu^{j-1}) - \pa_\mu G_j \right]^2  
\right) \, .
\eeq

We now compare this action to that of a latticized 5D warped gauge
theory\footnote{%
We choose an equally-spaced discretization of the coordinate $x_5$. 
It is possible to choose a different  latticization, for instance by uniformly discretizing in the conformal coordinates. This would match a different, non-equivalent deconstruction model with the same low-energy limit.}.
 We divide the fifth dimension into $N$ intervals of size $\Delta = \Lambda^{-1}$ (called the lattice spacing) and thus introduce $N+1$ lattice points $y_j$ ($y_0$ corresponds to $x_5 = 0$, while  $y_{N}$ corresponds to $x_5 = L$).
We thus  rewrite the continuum action \erefn{wga} as 
\beq
\label{e.5Dla}
S_5 \to  
 \int d^4 x  {\Delta  \over g_5^2} \left\{
- {1 \over 4}   \sum_{j = 0}^{N} F_{\mu\nu}(y_j) F_{\mu\nu}(y_j) 
 + {1 \over 2 }   \sum_{j = 1}^{N} a^2(y_j) 
 \left ({A_\mu(y_{j}) - A_\mu(y_{j-1}) \over \Delta}  - \pa_\mu A_5(y_j) \right )^2  
\right \} \, .
\eeq
It is natural to identify the inverse lattice spacing with the cutoff scale $\Lambda$ of the continuum theory. 
Comparing the 5D latticized action \eref{5Dla} with that of
 deconstruction, \eref{dqa}, we are able to set up the {\it
 dictionary} between 5D warped gauge theories and deconstructed
 models:
\bea
\label{e.dic}
y_j \Lambda  &\leftrightarrow& j \, ,
\nn 
 A_\mu(y_j) &\leftrightarrow& A_\mu^j \, , 
\nn
a(y_j)  A_5(y_j) &\leftrightarrow& g G_j \, ,
\nn 
a(y_j)\Lambda  &\leftrightarrow & g v_j \, ,
 \nn 
{g_5  \Lambda^{1/2}} & \leftrightarrow & g \, .
\eea
Of course this dictionary is ambiguous at higher orders in the lattice spacing $\Delta$,  as the latticization procedure is not uniquely defined (for example, discretizing $\pa_5$ is ambiguous).    

Specializing to the $\ads_5$ background  we choose the link vev's as 
\beq
 v_j = v q^{j} \, ,
\eeq 
with $q \leq 1$ for definiteness.\footnote{%
See ref. \cite{caergl} for a discussion of the  circumstances under which such a pattern can arise dynamically.} 
This leads to  the following $\ads_5$--deconstruction  dictionary  
\bea
\label{e.dick}
{k \over \Lambda}  & \leftrightarrow & \log q^{-1} \, ,
\nn
a(y_j)  & \leftrightarrow & q^j \qquad \qquad  \Rightarrow \qquad  \qquad  a_L  \leftrightarrow q^N  \, ,
\nn
\Lambda  & \leftrightarrow & g v \, ,
\nn
g_5^2  & \leftrightarrow & g/v \, .
\eea 
The UV brane corresponds to the site $j = 0$, and the IR one to $j =
N$. UV and IR brane matter fields can be represented in deconstruction
as matter fields transforming under the $0$-th or $N$-th group,
respectively.   

One often considers a formal limit of sending the lattice spacing to zero, that is  $\Lambda \to  \infty$.  
This is called the {\it continuum limit\/}. 
In deconstruction, this limit corresponds to $q \to 1$, $N \to \infty$, $v \to \infty$ with 
$q^N$ and $g v \log q^{-1}$ kept fixed.   
Strictly speaking, perturbative deconstruction can never reach the continuum limit, 
which is just a rephrasing of the triviality  problem of 5D gauge theories.
Indeed, from the last relation in \eref{dick} this would require either $g \to \infty$ or $g_5 \to 0$. 
For this reason we will refer to the continuum limit in the
following sense: 
on the 5D side, the {\it continuum theory} means the 5D theory with large enough  $\Lambda$  to neglect cutoff effects, in particular $\Lambda \gg k$, but still  $\Lambda \simlt 1/g_5^2$.
On the deconstruction side the continuum limit amounts to choosing $1 - q \ll 1$, $N (1 - q) > 1$.

Previous analytical studies in deconstruction were restricted to the case $q \ll 1$, far away from the continuum limit. 
As can be seen from \eref{dick}, this corresponds to the 5D theory with  the cutoff  smaller than the curvature scale. 
Such field theory is UV-sensitive as higher dimensional operators in the 5D action, 
e.g.  ${R \over \Lambda^2} F_{M N}^2$,  may be sizable.  
Thus, although deconstructed models with $q \ll 1$ are perfectly well-defined and interesting in their own right, their relation with  5D physics is obscure. 
In this paper we extend analytical studies to arbitrary $q$, including
$q \sim 1$. 
We will thus be able to approach deconstruction of 5D theories with $
k \ll \Lambda$, which are well under control on the 5D side.

\subsection{Kaluza--Klein approach}

We now turn to computing the spectrum of the theory. 
The gauge boson mass terms  are the following
\beq
\cl_{\mathrm mass} = \sum_{j = 1}^{N}  {1 \over 2} v^2 q^{2 j} (A_\mu^{j} - A_\mu^{j-1})^2 \, .
 \eeq
We perform the rotation $A_\mu^j = f_{j,n} A_\mu^n$, which brings the mass terms to the diagonal form   
$\cl_{\mathrm mass} = \sum_{n = 0}^{N}  {1 \over 2} m_n^2 (A_\mu^n)^2$.  
The coefficients $f_{j,n}$ should be viewed as a discretized KK profile $f_n(y_j)$. 
They satisfy the following difference equations (we define $x_n = m_n/g v$)
\beq
\label{e.de}
(q + q^{-1} - q^{-1} (x_n q^{-j})^2) f_{j,n}  - q f_{j+1,n} - q^{-1} f_{j-1,n} = 0
\eeq
subject to the boundary conditions
\beq
\label{e.dbc}
f_{0,n} = f_{-1,n} \, , \qquad f_{N,n} = f_{N+1,n} \, . 
\eeq
The normalization condition reads $g^{-2} \sum_{j=0}^N f_{j,n}^2 = 1$. 
Using the dictionary,  we can show that the difference equation \erefn{de} translates to the continuum equation for the KK profile, cf. \eref{wfeom}. 
The boundary conditions are obviously the discretized version of the Neumann boundary conditions, cf. 
\eref{wfbc}.

It is easy to find the zero-mode solution to eqs. \erefn{de} and \erefn{dbc}:
\beq
x_n = 0 \qquad f_{j,0} = {g \over \sqrt{N+1}} \, . 
\eeq
The wave function of the massive modes can also be found analytically.
We first define the variable $t[j] = x_n q^{-j}$ and the function  $F(t[j]) = q^j f_{j,n}$.
In terms of these variables \eref{de} becomes a q-difference equation
\beq
\label{e.qde}
(q + q^{-1} - q^{-1} t^2) F(t) - F(t q^{-1}) - F(t q) = 0 \, .
\eeq
This equation has been extensively studied in the mathematical
literature \cite{HE,Sthesis,kost,qNeumann,qEuclidean} and its solutions are called q-Bessel functions.  
Since, perhaps, readers are not so well acquainted with these
functions, we summarize all their relevant properties in \aref{qb}.  
The q-Bessel function $J_\nu(t;q^2)$ is defined by the series \erefn{qbd}. 
The other independent solution  $Y_\nu(t;q^2)$ is called the q-Neumann
function and is defined in \eref{qnd}.    
The q-Bessel functions generalize the ordinary (henceforth referred to as
{\it continuum}) Bessel functions and enjoy  similar properties (such as
integral representations, recurrence relations).   
In fact, for $q \to 1$ they are simply related to the  continuum ones,
see \eref{contJ}.  

Equation \erefn{qde} is a special case of the Hahn-Exton equation
\erefn{hee} with $\nu =1$ 
and the solution is $F(z) = A J_1(t;q^2) + B Y_1(t;q^2)$. 
Therefore the deconstructed KK profile can be written as\footnote{%
We could exactly
calculate the normalization of the KK profiles using results about
q-integrals~\cite{kost}, but here we simply note that the correct
normalizations are automatically included in the propagators in the next subsection,
and can be obtained from them.} 
\beq
\label{e.qkkp}
f_{j,n} = A_n q^{-j} \left [ 
Y_0(x_n;q^2) J_1(x_n q^{-j};q^2)  -   J_0(x_n;q^2) Y_1(x_n q^{-j};q^2) \right ] \, .
\eeq
The ratio of the two integration constants  has been chosen such that the first of the boundary conditions in \eref{dbc} is satisfied (the recursion relation \erefn{qbrr} is useful to prove this). 
The other boundary condition determines the KK mass spectrum. 
We find the spectrum  is given by solutions of the equation 
\beq
\label{e.qqc}
J_0(x_n;q^2) Y_0(q^{-N-1} x_n;q^2) - Y_0(x_n;q^2) J_0(q^{-N-1}x_n;q^2) = 0 \, . 
\eeq
Note the tantalizing formal similarity of eqs.\ (\ref{e.qkkp}),
\erefn{qqc} to the continuum KK profiles \erefn{kkp} and the
quantization condition \erefn{gbqc}. 
In the next section we will specify the parameter range in which the
continuum and deconstructed physics indeed match.

\subsection{Position-space approach}

Much as the continuum theory, deconstruction admits a
position-space picture.   
The deconstructed position propagator for the gauge field will be
denoted by  $P_{\mu\nu}^{jk}(p^2)$. 
The indices $j,k$ are obvious analogues of the position  variables
$x_5,y_5$ in the continuum propagator. 
Choosing the gauge--fixing term as 
\beq
\cl = - {1 \over 2 g^2 \xi} \sum_{j = 0}^N \left ( \pa_\mu A_\mu^j + \xi g^2 v (q^j G_j - q^{j+1} G_{j+1}) \right )^2     
\eeq
removes the mixing between $G_j$ and $A_\mu^j$ (in the above $G_0 \equiv 0 \equiv G_{N}$ is understood). 
In this gauge the propagator satisfies
$D_{\mu\rho}^{jm} P_{\rho\nu}^{mk} = i \delta^{jk}\eta_{\mu \nu}$,
 where $D_{\mu\nu}^{jk}$ is the kinetic operator (Fourier-transformed to 4D momentum space): 
\beq
g^2 D_{\mu\nu}^{jk} = 
\left (-p^2 \eta_{\mu\nu} + (1 - 1/\xi) p_\mu p_\nu \right )\delta_{jk}   
+  g^2 v^2 q^{2j} \left (  \delta_{jk} + q^2 \delta_{jk} -
\delta_{j,k+1} - q^2\delta_{j,k - 1} \right )  \eta_{\mu\nu} \, .
\eeq
The propagator is of the form 
$P_{\mu\nu}^{jk}(p^2) = (\eta_{\mu\nu} - {p_\mu p_\nu \over p^2} ) P_{j,k}(p^2) + {p_\mu p_\nu \over p^2} P_{j,k}(p^2/\xi) $, 
where $P_{jk}(p^2)$ satisfies   
\beq
\label{e.ppaux1}
\left (-p^2  + g^2 v^2 q^{2j} (1 +  q^2) \right ) P_{j,k}  - g^2 v^2
q^{2j} (P_{j-1,k} + q^2  P_{j+1,k}) = i g^2 \delta_{j,k} \, ,
\eeq
together with the boundary conditions 
\beq
\label{e.ppnbc}
P_{0,k} = P_{-1,k} \, , \qquad P_{N,k} = P_{N+1,k} \, .
\eeq
These boundary conditions, technically speaking,  are chosen such that \eref{ppaux1} is correct for $j= 0$ and $j= N$. 
They are analogues of the Neumann boundary conditions for the continuum propagator.
See \aref{dbc} for the deconstruction of general boundary conditions. 

For $j \neq k$, the propagator equation \erefn{ppaux1} is the same as
the KK mode equation \erefn{de} with   $x_n \to x \equiv p/g v$. 
Therefore we can  easily write down the solutions for $j < k$ and $j > k$ that satisfy  the relevant boundary condition: 
\bea
\label{e.ppaux2}
P_{j,k}^< 
&=&
 A_k^< q^{-j} \left [ Y_0(x;q^2)  J_1(q^{-j} x;q^2) -  J_0(x;q^2)  Y_1(q^{-j} x;q^2) \right ] 
\nn
 P_{j,k}^>
 &=&
 A_k^> q^{-j} \left [ Y_0(q^{-N-1} x;q^2)  J_1(q^{-j} x;q^2) -  J_0(q^{-N-1} x;q^2)  Y_1(q^{-j} x;q^2) \right ]  \,.
\eea  
The constants $A_k^{<,>}$ are determined by the matching conditions 
\beq
P_{k,k}^< = P_{k,k}^> \, , \qquad P_{k+1,k}^< = P_{k+1,k}^> + {i \over
  v^2 q^{2k+2}} \, ,  
\eeq
which make the solution  \erefn{ppaux2} satisfy the propagator equation for $j = k$ and $j = k+1$. 
After some algebra we solve for $A_k^{<,>}$ and derive the deconstructed position  propagator:
\beq
\label{e.dpp}
\ba{l} 
\ds
P_{jk}(p^2) = {i \pi \over  v^2 (1 - q^2)} q^{-k-j} \, \left[Y_0(x;q^2)
  J_1({q^{-j} x}  ;q^2) -  J_0(x;q^2)  Y_1({q^{-j} x};q^2) \right] 
\\ \ds
\frac{\left[Y_0(q^{-N-1} x;q^2)  J_1({q^{-k} x}  ;q^2) -  J_0(q^{-N-1}
  x;q^2)  Y_1({q^{-k} x};q^2)\right] } { Y_0(x;q^2) J_0(q^{-N-1} x;q^2)
  -J_0(x;q^2) Y_0(q^{-N-1} x;q^2) } \, ,
\ea
\eeq
for $j \leq k$. 
The denominator reproduces the poles at $p^2$ equal to the KK masses $m_n^2$ given by \eref{qqc}. 

We can also define ``brane-to-brane'' propagators
\bea
\label{e.dbbpp}
P_{00}(p^2) &=& 
{i g \over  v p} {Y_0(q^{-N-1} x;q^2)  J_1( x  ;q^2) -  J_0(q^{-N-1} x;q^2)  Y_1(x;q^2) 
\over
 Y_0(x;q^2) J_0(q^{-N-1} x;q^2) -J_0(x;q^2) Y_0(q^{-N-1} x;q^2) } \, ,
\nn
P_{NN}(p^2) &=& 
{i g \over q^{N} v p}  {Y_0(x;q^2)  J_1({q^{-N} x} ;q^2) -  J_0(x;q^2)  Y_1({ q^{-N} x};q^2) 
\over 
Y_0(x;q^2) J_0(q^{-N-1} x;q^2) -J_0(x;q^2) Y_0(q^{-N-1} x;q^2) } \, ,
\nn
P_{0N}(p^2) &=& 
{i g^2  (1 - q^2) \over  \pi p^2} {1 \over  Y_0(x;q^2) J_0(q^{-N-1}
  x;q^2) -J_0(x;q^2) Y_0(q^{-N-1} x;q^2) } \, .
\eea
We will also need the UV propagator evaluated at Euclidean momenta $p^2 = -p_E^2$.
It is convenient to rewrite it in terms of the modified q-Bessel functions $I_\nu(x;q^2)$ and $K_\nu(x;q^2)$ defined in eqs. \erefn{mqbfk} and \erefn{mqbsk}: 
\beq
\label{e.qbeuvp}
P_{00}(- p_E^2) = 
{i g \over v p_E} {
K_0(q^{-N-1} x_E;q^2)  I_1(x_E;q^2) +  I_0(q^{-N-1} x_E;q^2)  K_1(x_E;q^2) 
\over 
K_0(x_E;q^2) I_0(q^{-N-1} x_E ;q^2) - I_0(x_E;q^2) K_0(q^{-N-1}
x_E;q^2) } \, ,
\eeq 
where $x_E = p_E/gv$.

The connection between the KK and position pictures in deconstruction is provided by the spectral formula 
\beq
P_{jk}(p^2) = - i \, \sum_{n=0}^N \frac{f_{j,n} f_{k,n}} {p^2 - m_n^2} \, ,  
\label{e.spectraldec} 
\eeq
which involves a finite sum. 
This implies that the propagator \erefn{dpp} is, in fact, given by a ratio of two polynomials in $p^2$.   
 

\subsection{Holographic approach}

Finally, it is possible to implement the holographic approach in deconstruction.
In fact, at the conceptual level, constructing  the boundary effective action is even clearer here.
The procedure in deconstruction amounts to integrating out $N$ gauge
bosons $A_\mu^j$ with $j\geq 1$ and leaving $A_\mu^0$ as the low
energy variable.\footnote{%
Note that neither ${A}_\mu^0$ nor the remaining gauge bosons are mass eigenstates.
However, integrating out linear combinations of light and massive fields  is equivalent to integrating out heavy mass eigenstates after appropriate redefinitions of the light fields in the low-energy theory ~\cite{baporast}.}
We obtain, for $\xi = 1$, 
\beq
\label{e.dh}
S_{\mathrm{eff}} = - \frac{1}{2} \int {d^4 p \over (2\pi)^4}
 {A}^0_\mu(p) \Pi_{\mu \nu} {A}_\mu^0(p) \, , \qquad 
 \Pi_{\mu \nu} = -{i \over P_{00}(p^2)} \eta_{\mu\nu} \, .
\eeq

\section{Discussion}
\label{s.di}

In this section we give a detailed discussion of  the spectrum, the KK profiles and the propagators derived in \sref{de}.
We first consider the case $q \sim 1$ and determine the energy range in which deconstruction is a good approximation of  the continuum 5D gauge theory. 
Then, we extend our discussion to general $q$, including the case $q \ll 1$, which does not have a 5D interpretation. 
Finally, we comment on the holographic interpretation of the deconstructed setup, 
for both  the continuum case $q \sim 1$ and for smaller values of $q$.

Let $q \sim 1$, but still $q^N \approx q^{N+1} \ll  1$. 
In this case it is convenient to rewrite $q = 1 - \delta$ with
$\delta$ (but not $N \delta$) being a small parameter.  
From \eref{dick} $\delta$ is interpreted as the ratio $k/\Lambda$, 
that is $ k \leftrightarrow \delta g v$ up to corrections of higher
order in $\delta$. 
From this it follows that  the smallness of $\delta$ is a necessary condition for the  corresponding 5D theory to be in a controllable regime. 

We first discuss the KK spectrum of deconstruction for $\delta \ll 1$. 
For small enough $\delta$, according to  \eref{contJ}, q-Bessel
functions can be approximated by the continuum ones, as long as their
argument is less than $1$. 
Thus, for $m_n \simlt q^N g v$, the eigenvalue equation \erefn{qqc} can be approximated as  
\beq
\label{e.dqca}
J_0 \left ({m_n \over \delta g v} \right) Y_0 \left ({m_n \over \delta g v q^{N}} \right) 
- Y_0 \left ({m_n \over \delta g v} \right) J_0 \left ({m_n \over \delta g v q^{N}} \right) 
 \approx 0 \, .
\eeq 
The spectrum obtained by solving  \eref{dqca} matches the continuum spectrum with 
$k  =  \delta g v$ and $a_L = q^N$, as prescribed by the dictionary \erefn{dick}. 
In particular, it is linearly spaced: 
\beq
m_n \approx  \pi \delta g v q^{N} (n - \frac{1}{4}) \, , \qquad \qquad \qquad
m_n \simlt q^{N} g v \, . 
\eeq
One can also easily see that the deconstructed position propagators match the continuum ones  (more precisely, $P(p^2,y^j,y^k) \approx P_{jk}(p^2)$ after relating the parameters as in \eref{dick}) for momenta smaller  than $q^{N} g v$. 
This ensures that both theories describe the same physics  below the scale $q^{N+1} g v$ at leading  order in $\delta$ and $p/g v$. 

For $m_n > q^{N} g v$, some of the q-Bessel functions  describing the KK
profile and the mass spectrum fall outside the range of continuum
approximation.  
In particular, the functions $Y_0( m_n/ q^{N+1} g v;q^2)$ and
$J_0(m_n/ q^{N+1} g v;q^2)$ in the quantization condition \erefn{qqc}
can no longer be 
approximated by continuum Bessel functions.  
According to \eref{qbltl} these functions  oscillate rapidly  with
exponentially  decaying/growing amplitudes.  
On the other hand, for  $m_n \ll \delta g v$,  we have
$|Y_0({m_n/g v};q^2)| \gg |J_0({m_n/g v};q^2)|$.  
We can thus infer that the  solutions to \eref{qqc} for 
$ q^N g v \simlt m_n \simlt \delta g v $  are  approximately given by
the asymptotic zeros of  $J_0({m_n/q^{N+1}  g  v};q^2)$. 
Using,  \eref{qbltlz} we find an {\it exponential spectrum}, $m_n \approx g v q^{N+1} q^{-n}$,  in this
regime. 
This formula can be refined using the detailed asymptotic
behaviour of the q-Bessel functions at small and large argument, given
by Eqs.~\refeq{e.smallarg}, \refeq{e.qbltl} and \refeq{e.qYltl}. We
obtain\footnote{Note that for weak warping,
$q^{N+1} \simlt  \delta$, there is no regime where this approximation
is valid.}
\beq
m_n \approx g v q^{N+1} q^{-n+\alpha_n}  \, ,  \qquad \qquad \qquad 
q^{N+1} g v \simlt m_n \simlt \delta g v \, , \label{e.heavyspectrum}
\eeq
where
\beq
\label{e.qalpha}
q^{\alpha_n} =
\left(1-\frac{1}{N+2-n+\frac{\gamma_{q^2}-\log(1-q^2)}{\log 
    q}} \right)^{-\frac{1}{2}} \, .
\eeq
The q-Euler--Mascheroni constant $\gamma_{q^2}$ is defined in
\erefn{qeulermascheroni}. The upper limit in the region of validity
of~\refeq{e.heavyspectrum} arises from the use of
the leading terms in the expansions of the q-Bessel functions at small
argument. Moreover, in order to be able to use \refeq{e.qYltl}, we
have made the ansatz $|\alpha_n|\ll 1$, which is indeed satisfied
by~\refeq{e.qalpha} when $m_n \simlt \delta g v$. 

Closer to the top of the KK tower, for $\delta g v \simlt m_n \simlt
2gv$, 
we have not
found an accurate analytical formula for the spectrum.  
But from the asymptotic form of the q-Bessel functions
and the trend for the masses below
$\delta g v$, it can be argued that a growth  stronger
than exponential 
will persist until the scale $\sim 2 g v$. This agrees with an
extrapolation from the known results at small~$q$ and with the exact
formula for the product of non-vanishing eigenmasses in Ref.~\cite{kash}.
The numerical computations confirm this behaviour. 
For $m_n > 2 g v$, there is no solution to the quantization condition:
the deconstructed KK tower is cut off. 
\begin{figure}[t]
\begin{center}
\centerline{\includegraphics[width=0.7\textwidth]{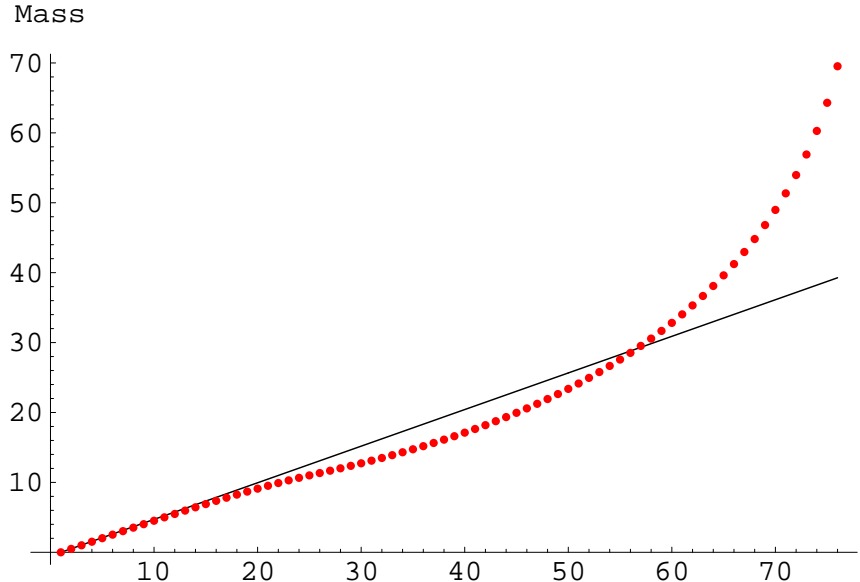}}
\caption{{\footnotesize 
Numerical comparison of the deconstructed KK spectrum (red circles)
and the continuum one (black line).  KK masses are given in units of
$k$.
The deconstructed spectrum is calculated for $N = 75$ and $q = 0.97$,
which corresponds to the warp factor $a_L \approx 0.102$ on the
continuum side. 
\label{fig.contvsdec} 
}}   
\end{center}
\end{figure}

Summarizing, we see that above the scale $q^N gv$, the spectrum
depends (approximately) exponentially on $n$, rather than linearly, and
resembles nothing one encounters in a 5D theory.\footnote{%
The exponential spectrum~\refeq{e.heavyspectrum} is a specific feature of the deconstructed model considered here.
We have checked that the spectrum outside the continuum regime is very
different in the deconstructed model corresponding to a uniform
discretization in the conformal coordinates.}
It is worthwhile to emphasize that
the linear ``continuum'' regime is not just the linear approximation of the subsequent exponential regime, at least for small $\delta$. 
In fact, for the first few modes in the exponential regime, the masses
grow more slowly than in the linear regime. This can be clearly seen
in Fig.~\ref{fig.contvsdec}. 

\begin{figure}[t]
\begin{center}
\centerline{\includegraphics[width=0.8\textwidth]{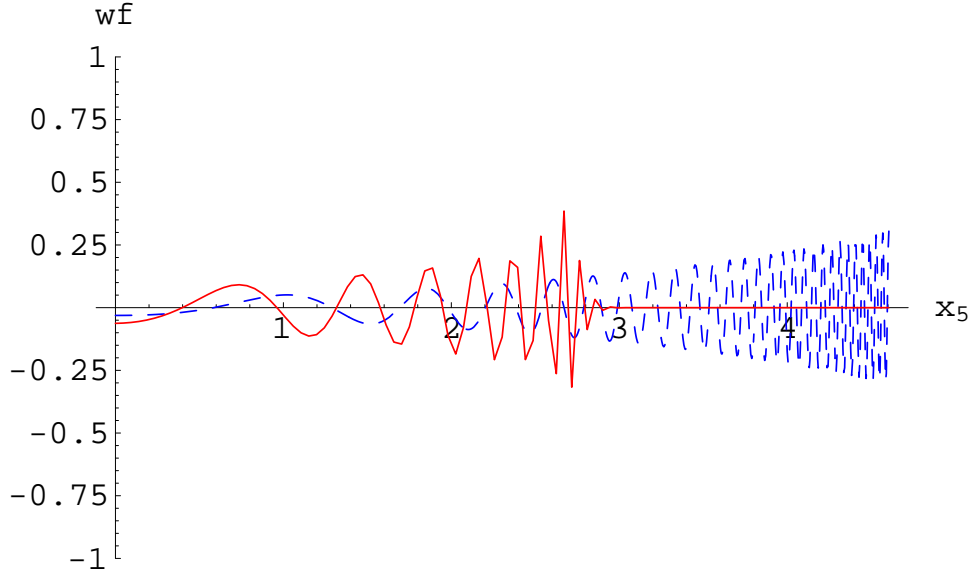}}
\caption{{\footnotesize 
Deconstructed (red continuous line) and continuum (blue dashed line)
KK profiles of the mode $n=55$, for  $N = 100$ and $q = 0.955$,
which corresponds to the warp factor $a_L \approx
0.010$. \label{fig.wavefunction} }}   
\end{center}
\end{figure}

A similar  analysis can be applied to study  the KK profiles. 
As we have already mentioned, for 
KK modes with masses below $g v q^N$ they match the continuum ones. 
Above this scale, the deconstructed KK profiles are still described by
continuum Bessel functions for $j < j_*$, where $ q^{j_*} = m_n/gv$.   
However, since the mass of the deconstructed $n$-th mode is different from the
continuum one, the deconstructed profile is different as well.   
The normalization factor is different too, because of the behaviour at large
$j$.  For $j > j_*$, the quantization
condition~\refeq{e.qqc} and the asymptotic behaviour of the
q-Bessel functions given by eqs.\ \refeq{e.genas}, \refeq{e.qbltl} and
\refeq{e.qYltl} imply that the 
deconstructed KK profiles are exponentially 
damped, in contrast with the continuum ones, which oscillate with
increasing amplitude. 
Hence, above $g v q^N$, deconstructed KK modes are decoupled from the
IR brane, while they are more strongly coupled than the continuum ones
to the UV brane,  because of a larger normalization factor. We plot in
Fig.~\ref{fig.wavefunction} a typical deconstructed KK profile in the
regime $m_n \gg gv q^N$, together with its continuum counterpart.

The preceding discussion identified the {\it deviation scale} 
\beq
\Lambda_D = g \, v  \, q^{N} \, ,
\eeq
above which the deconstructed KK spectrum and profiles no longer match those of the continuum theory.
Using the dictionary \erefn{dick} the deviation scale is translated to
$\Lambda a_L = \Lambda(L)$, which is  the cutoff scale on the IR
brane.  
Note that for $q \sim 1$ this scale  is much larger than the KK scale
$M_{KK} \sim \delta  g v q^{N}$ and there are  $\sim 1/\delta$ KK
excitations that are properly matched. 
  
However the applicability range of the continuum theory is far larger,
as long as we keep to  the appropriate observables. 
For example, as we discussed above,  5D UV brane physics remains
perturbative up to a much higher scale, of order $\Lambda_S(0) = 1/g_5^2$.  
Naively, it would seem that deconstruction cannot reproduce continuum
UV brane physics above the scale $\Lambda_D$. We will show however that it
does.

Consider the deconstructed UV propagator at Euclidean momenta, \eref{qbeuvp}. 
According to~\refeq{e.qIltl} and~\refeq{e.qKltl}, the modified
q-Bessel functions 
\eref{qbeuvp} share the property with 
their continuum analogues that at large argument $K_\nu(x;q^2) \ll
I_\nu(x;q^2)$, including the $x \gg 1$ region where they deviate from
the continuum $K$ and $I$ functions.  
Thus, for $p_E \simgt  \delta g v q^N$ we can approximate     
\beq
P_{00}(-p_E^2) \approx {i g \over v p_E} 
{K_1({p_E \over g v};q^2) 
\over 
K_0({p_E \over g v};q^2)} \approx 
{i g \over  v p_E} 
{K_1({p_E \over \delta g v}) 
\over 
K_0({p_E \over \delta g v})} \, .
\eeq
The last equality involving only continuum Bessel functions holds for
$p_E \simlt g v$. 
We can see that $P_{00}$ correctly reproduces the continuum UV propagator 
(with $k = \delta g v$, $g_5^2 = g/v$) at energies below $g v$,
cf.\erefn{bbpw}  and \erefn{bbpf}.  
This ensures that the two frameworks describe the same UV brane
physics (for example, scattering amplitudes of UV brane localized
matter fields) below  $g v$.   
The limiting scale is translated to $\Lambda = \Lambda(0)$ -- the
cutoff scale on the UV brane.  

\begin{figure}[t]
\begin{center}
\centerline{\includegraphics[width=0.5\textwidth]{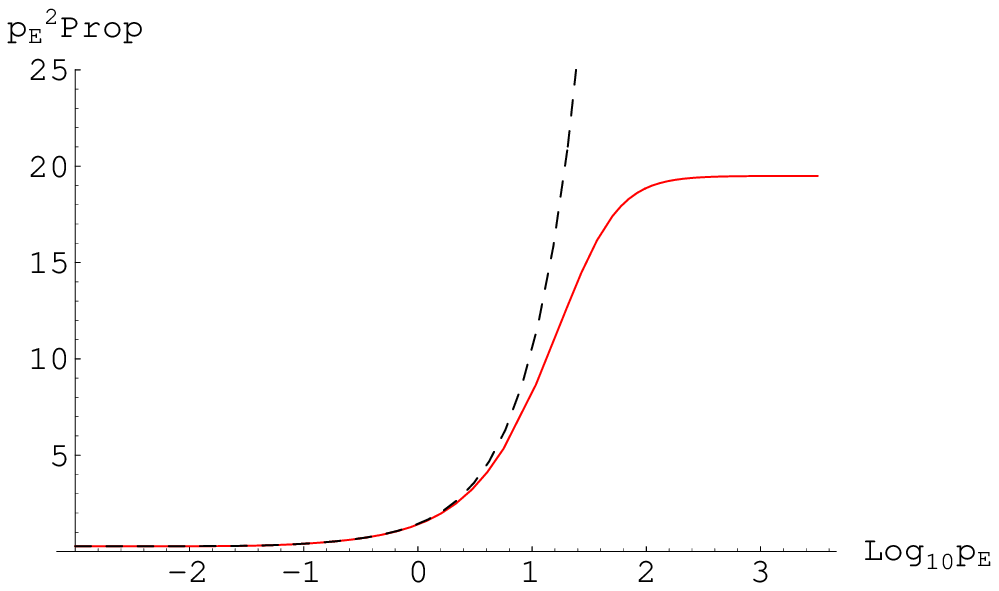}
\includegraphics[width=0.5\textwidth]{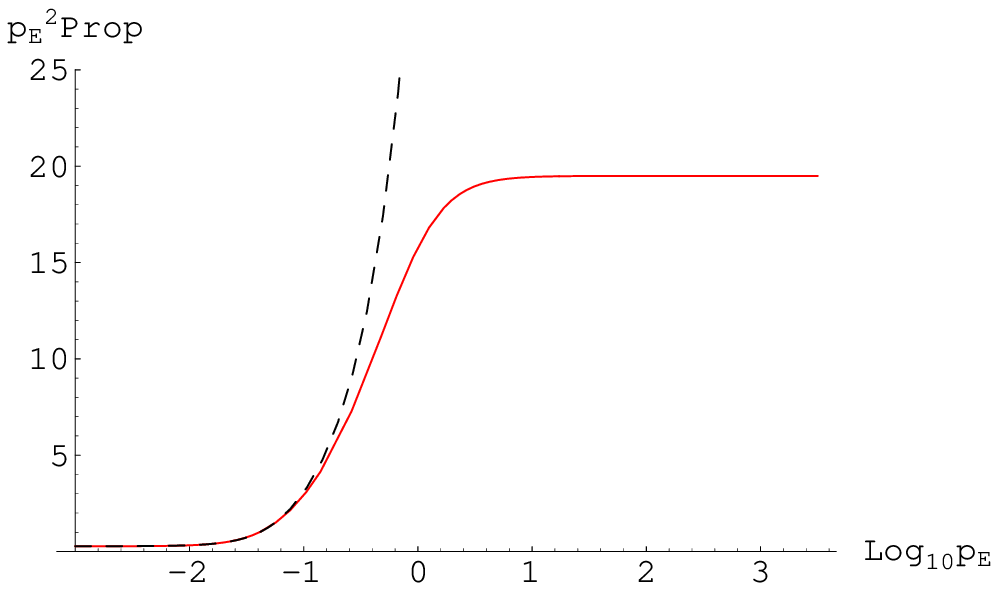}} 
\caption{{\footnotesize 
Deconstructed (red continuous line) and continuum (black dashed
line) Euclidean brane-to-brane propagators at the UV (left) and IR (right)
branes, for $N=70$ and $q=0.95$, and thus $a_L\approx 0.0276$. Momenta 
are given in units of $k$. 
\label{fig.props} }} 
\end{center}
\end{figure}

On the other hand, on the IR brane the matching is terminated at a lower scale. 
For $p_E \simgt g v q^N$ the deconstructed propagator is purely four-dimensional, 
$P_{NN}(-p_E^2) \approx i g^2 /p_E^2$, and deviates from  the $1/p_E$
behaviour of the continuum IR propagator. Thus IR physics is matched by
deconstruction only below the scale  $g v q^N$, which is translated to
the cutoff scale on the IR brane  $\Lambda(L)$.  
 We compare the continuum and
deconstructed UV and IR propagators in Fig.~\ref{fig.props}.

This discussion can be easily generalized to show that,  for an
arbitrary position $y$ in the bulk, deconstruction reproduces continuum
physics up to $\Lambda(y) = \Lambda a(y)$.  
Therefore, the deconstruction framework we consider  is indeed a
realization of a position-dependent cutoff.

Let us pause for a moment to summarize what we have shown so far. 
Our results imply that deconstruction provides a correct approximation of the continuum physics at the UV and IR branes, and at any position in the bulk.   
The matching holds all the way up to the respective position-dependent cutoff scale.   
In particular, for $\Lambda \sim 1/g_5^2$ deconstruction works well
all the way up to the position-dependent strong-coupling scale, that
is in the entire perturbativity range of the continuum theory.\footnote{%
On the other hand, if we choose $\Lambda$ sufficiently smaller than $1/g_5^2$, 
the 4D behaviour of the deconstructed theory sets in early enough to
stop the linear growth of tree-level amplitudes and prevent the
occurrence of strong coupling.  
Therefore, this scenario can  provide, in principle, a perturbative completion of the continuum $\ads_5$  theory that can be valid up to much higher scales.} 
The correspondence extends to the energy range where the KK spectra and profiles in the two theories are completely different.   
 
We were able to establish the agreement of the deconstructed and continuum
theories in the position approach.  
However, it all seems miraculous from the point of view of the KK approach. 
As we have discussed before,  the deconstructed KK masses and profiles
deviate from 
the continuum one at the scale $\Lambda_D$.  
Thus, in the two frameworks scattering amplitudes above $\Lambda_D$
involve an exchange of a different number of KK modes with different
masses and couplings to the brane fields. 
Somehow, for UV brane amplitudes the two effects cancel and both
frameworks predict the same result. The reason for the better
agreement of the UV propagators is the locality of the
action~\refeq{e.dqa} in the extra-dimensional lattice: the Euclidean
UV propagator is screened from the IR brane. On the
other hand, the KK masses are determined by global properties of the
theory and are thus sensitive to the physics at the IR brane. 

Of course, strictly speaking, the matching of the propagators holds
only for $t$-channel amplitudes, when the propagators carry space-like
momenta. An $s$-channel experiment on the UV brane with enough precision
could resolve individual KK resonances, which do not fit in the two
theories, and distinguish our completion from a strongly-coupled
continuum theory.
In this case, above $\Lambda_D$, the matching holds only when the
energy spread of the scattered beams is sufficiently larger than the
KK spacing, so that individual resonances cannot be resolved.

We now move to discussing the situation when the parameter $q$ is not
close to $1$, that is when deconstruction is away from the continuum
limit. 
In this case there is no energy regime where the deconstructed
spectrum mimics the continuum one.
However, earlier works that concentrated on $q \ll 1$ showed that also
in this regime deconstruction reproduces certain features of the
continuum theory, for example logarithmic running of gauge couplings. 
We are now in a position to make the relation more precise and
interpolate between the different values of $q$. 

\begin{figure}[t]
\begin{center}
\centerline{\includegraphics[width=0.7\textwidth]{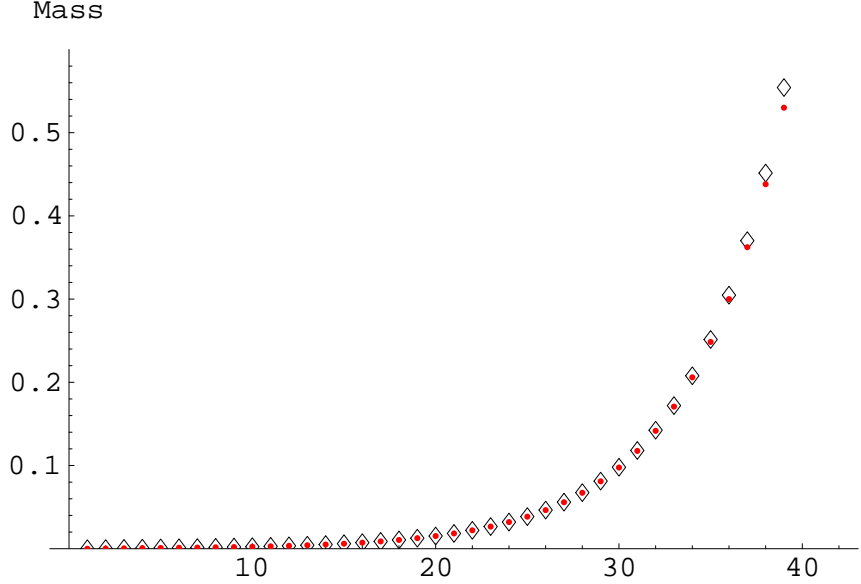}} 
\caption{{\footnotesize 
Comparison of our approximate analytical spectrum (black empty
diamonds) and the numerical one (red solid circles) for $N = 50$ and
$q = 0.832$, and thus $a_L\approx 10^{-4}$, in units $k=1$. We plot
only the first 40 modes, which lie in the region of validty of
our approximation. The relative error is below 1\% for all but the
last few modes. 
\label{fig.comparespectrum} }} 
\end{center}
\end{figure}

At the level of the spectrum, for general $q$ we can again distinguish
three regimes. For $m_n \simlt q^{N+1} g v$, the spectrum is also linear, but
now this regime holds only for a few modes. For $m_n \simgt q^{N+1} g v$, the KK
masses are given with great precision by~\refeq{e.heavyspectrum}, and
only the last few modes with $(1-q) g v \simlt m_n \simlt g v$ deviate;
see Fig.~\ref{fig.comparespectrum}. In the limit $ q \ll 1$ we see that
only the exponential regime survives, in agreement
with~\cite{faki,rashwe,kash}. The behaviour of the KK profiles is
qualitatively the same as the one discussed before.

Consider now, once more, the UV propagator \erefn{qbeuvp}, but now
for arbitrary $q$.  
At large argument $I$ still dominates over $K$, so that 
at energies above the KK mass gap, $p_E \simgt g v q^{N+1}$, we can approximate    
\beq
P_{00}(- p_E^2) \approx 
{i g \over  v p_E} { 
K_1(p_E/gv;q^2)  \over  K_0(p_E/gv;q^2) } \, .
\eeq  
As long as $p_E \simlt g v$, we can use the small argument
asymptotics \erefn{smallarg} of the modified q-Bessel functions to obtain 
\beq
\label{e.qbsquvp}
P_{00}(- p_E^2) \approx 
{1 \over \log \left ( gv(1-q^2)e^{-\gamma_{q^2}}/q^{1/2} p_E \right )} {i g^2 \log
  q^{-1} \over p_E^2} \, .
\eeq  
The deconstructed UV propagator exhibits  analogous momentum dependence to that of the continuum UV propagator below the curvature scale, cf. \eref{bbpw}.  
The factor $g_5^2 k$ in the continuum is replaced here by $g^2 \log q^{-1}$, in agreement with the dictionary \erefn{dick}.  
Away from the continuum limit, for $q$ much smaller than $1$, only the argument of the ``classical log'' does not match the continuum value.  
Therefore deconstruction with arbitrary $q$ captures the essentials of the continuum UV brane physics below the scale $k$.  
This explains why deconstruction could reproduce certain features of the 5D theory, such as the logarithmic running of gauge couplings \cite{running}, also in the parameter space away from the continuum limit.  
%

This brings us to a comment on the relation of deconstruction and holographic CFTs. 
We have argued that the UV boundary physics is reproduced in deconstruction for arbitrary $q$, below  the scale $g v$ where the approximation \erefn{qbsquvp} is valid.  
Therefore it is natural to conjecture that deconstructed models, also in the parameters range where they do not have a 5D interpretation, describe the dynamics of some large $N_c$ strongly-coupled theories which are approximately conformal over a  range of scales.
This is an interesting generalization of the AdS/CFT conjecture. 
In the case at hand 4D strongly-coupled theories might be dual to
weakly-coupled 4D theories with extended gauge symmetry.   
Matching the propagator \erefn{qbsquvp}  onto the quark bubble calculation in QCD at large Euclidean momentum, the number of colours in the CFT can be related to the parameters of deconstruction:
\beq 
N_c = {12 \pi^2 \over g^2 \log q^{-1}} \, .
\eeq
According to the dictionary \erefn{dick} we have $g_5^2 k \llra g^2 \log q^{-1}$,  
so this  a translation of the well-known  AdS/CFT relation  $N_c = 12 \pi^2/g_5^2 k$.


\section{Summary and outlook}

In this paper we clarified in what sense and in which parameters range deconstruction approximates 5D gauge theories in the Randall--Sundrum background. 
In our analysis of the deconstructed theory, we employed powerful tools of the mathematical theory of q-Bessel functions. 
This allowed us to study analytically the parameter space that was previously accessible only to numerical methods, including the $q \sim 1$ region where the corresponding 5D theory is under control. 

The main result of this paper is an explicit proof that a warped 5D gauge theory can be approximated by deconstruction in all its perturbativity range. 
More precisely, deconstruction provides a faithful description all the way up to the {\it position-dependent cutoff} $\Lambda(x_5) =  a(x_5)\Lambda $, where  $\Lambda$ is the inverse lattice spacing. 
In particular, the continuum theory with $\Lambda \sim  1/g_5^2$ can be reproduced by deconstructed models   with perturbative gauge coupling, $g \sim 1$, in which case the  matching extends all the way up to  the {\it position-dependent strong-coupling scale} $\Lambda_S(x_5) = a(x_5)/g_5^2$. 
This is not a trivial result, as the KK spectra of the two
theories deviate at a much lower scale, $\Lambda_D = a_L \Lambda$.

The technical results we derived can be readily applied to more phenomenological studies.
In fact, we have shown that computations in  deconstructed warped theories can be performed at the quantitative level comparable to that in 5D.   
This is an encouragement to study deconstructed versions  of phenomenological models in $\ads_5$, such as the models of the electroweak sector of refs. \cite{higgs}. 
Deconstruction could provide for a UV completion  of these models that could remain perturbative to much higher energy scales.   
Furthermore, deconstruction offers a playground to study the evolution
of gauge couplings in $\ads_5$ \cite{running,rasc}. 
This was already exploited in \cite{faki,rashwe}, but only in the region $q \ll 1$, where the link with the 5D computation is obscure.  
The virtue of deconstruction in this case it that it provides a concrete  physical realization of the 5D cutoff physics,  which allows, in particular, to study threshold effects.   

It would also be interesting to exploit the relation of deconstruction with holographic CFTs. 
Most interestingly, deconstruction can describe certain aspects of the real-world QCD in the strongly coupled regime.
For example, chiral symmetry breaking and physics of $\rho$ and $a$ resonances can be captured, 
as  shown  in ref. \cite{sost}.
More recently, effective description of low energy QCD was pursued in 5D continuum models \cite{adsqcd}. 
Our results allow for quantitative studies of deconstructed versions  of these models. 
This could shed further light on the origin of this AdS/QCD
correspondence and its connection to the Migdal's approach~\cite{Migdal} discussed recently  in~\cite{erkrlo}. 
Indeed, the high energy  behaviour of the continuum correlation functions, which is non-analytic in $p^2$,  
is approximated in deconstruction  by a ratio of polynomials, similarly as in the Migdal's approach.

Finally, we note that the methods used here can be directly applied to
fields with different spins, at least in the massless case. For
instance, the difference equation for the graviton corresponds to the
Hahn--Exton equation~\refeq{e.hee} with $\nu=2$. Therefore, all the results presented here for gauge bosons translate in a straightforward way to the gravity case. 
In particular, the qualitative features of the spectrum are the same, and agree with the discussion presented in ref.~\cite{rascth}.
It is also likely that the methods of q-difference equations can  be applied to study deconstruction of more general backgrounds than the Randall--Sundrum $\ads_5$ one.

\section*{Acknowledgements}
We are indebted to R.~F.~Swarttouw for sharing with us his expertise
on q-Bessel functions.  
We also thank A.~Pomarol and R.~Rattazzi for useful discussions, and
F.~del~Aguila and K.~y.~Oda for collaboration 
in the initial stages of this work.
A.F.\ and S.P.\ were partially supported by the European Community 
Contract MRTN-CT-2004-503369 for the years 2004--2008 and by 
 the MEiN grant 1 P03B 099 29 for the years 2005--2007.
J.B.\ and M.P.V.\ have been
partially supported by MEC (FPA 2003-09298-C02-01) and by
Junta de Andaluc{\'\i}a (FQM 101 and FQM 437). J.B.\ also thanks MEC
for an FPU grant.

\renewcommand{\thesection}{Appendix \Alph{section}} 
\renewcommand{\theequation}{\Alph{section}.\arabic{equation}} 
\setcounter{section}{0} 
\setcounter{equation}{0} 

\section{Implementation of general boundary conditions}
\label{a.dbc}

We generalize our results to gauge theories with Dirichlet boundary conditions on one or on both branes.  

The mixed momentum--position space propagator for a gauge boson in a slice of $\ads_5$ can be succinctly written as 
\beq
\label{e.ppgbc}
P^{\alpha \beta}(p^2,x_5,y_5)= {i \pi g_5^2 \over 2 k}
{ \left ( A^\alpha[0] J_1({p \over k a(x_5)}) -  B^\alpha[0] Y_1({p \over k a(x_5)}) \right )
\left ( A^\beta[L] J_1({p \over k a(y_5)}) - B^\beta[L]Y_1({p \over k a(y_5)}) \right )
\over a(x_5) a(y_5) \left (A^\alpha[0] B^\beta[L] - B^\alpha[0] A^\beta[L] \right )} \, ,
\eeq
where the parameters $A^\alpha[0]$, $B^\alpha[0]$, $A^\beta[L]$,
$B^\beta[L]$ depend on the boundary conditions.
They take the values 
\beq
A^N[x_5] = Y_0\left ({p \over k a(x_5)}\right) \, , \qquad \qquad 
B^N[x_5] = J_0\left({p \over k a(x_5)}\right) 
\eeq
if Neumann boundary conditions are imposed at $x_5$, and
\beq
A^D[x_5] = Y_1\left({p \over k a(x_5)}\right) \, ,\qquad \qquad 
B^D[x_5] = J_1\left({p \over k a(x_5)} \right ) 
\eeq
for Dirichlet boundary conditions at $x_5$.
We can also define brane-to-brane propagators, as long as we deal with Neumann boundary conditions on the given brane.  
For example, for the Neumann--Dirichlet case we obtain the UV brane-to-brane propagator
\beq
P_{\mathrm UV}^{\mathrm ND}(p^2) \equiv P^{\mathrm ND}(p^2,0,0)
=  {i g_5^2 \over p}
{Y_1({p \over k a_L})J_1({p \over k}) -  J_1({p \over k a_L})Y_1({p \over k})
\over
Y_0({p \over k})J_1({p \over k a_L}) -  J_0({p \over k})Y_1({p \over k a_L})} \, ,
\eeq
which is suppressed for $\sqrt{-p^2} \ll M_{\mathrm KK}$, while for $\sqrt{-p^2} \gg M_{\mathrm KK}$ it is indistinguishable from  the UV propagator in the Neumann--Neumann case.  

We turn to discussing the method of realizing these more general boundary conditions in deconstruction.
To this end we include two more scalar fields: 
$\Phi_0$  charged under $A_\mu^0$ with a vev $v_0/\sqrt{2}$ and  
$\Phi_{N+1}$ charged under $A_\mu^N$ with a vev $v_{N+1}/\sqrt{2}$, so that the boundary entries in the gauge boson mass matrix are modified. 
The effect of the additional vev's is to modify the  ``boundary conditions'' for the propagator. 
More precisely, \eref{ppaux1} represents  the correct propagator equation if we impose 
\bea
\label{e.dppmgbc}
P_{-1,k} & = & {v^2 -  v_0^2 \over v^2} P_{0,k} \, ,
\nn
P_{N+1,k} & = & {v^2 - q^{-2(N+1)} v_{N+1}^2 \over v^2} P_{N,k} \, .
\eea 
Setting $v_0 = v_{N+1} = 0$ we recover the Neumann boundary conditions \erefn{ppnbc}. 
With  $v_0 =  v$ we get $P_{-1,k} = 0$ which mimics the Dirichlet boundary conditions on the UV brane. Similarly $v_{N+1} = q^{N+1} v$ yields  $P_{N+1,k} = 0$, which is a deconstructed version of  the Dirichlet boundary conditions on the IR brane. 
Other choices of $v_0$ and $v_{N+1}$ correspond to mixed boundary conditions on the continuum side. 

We are now in a position to write the general form of the deconstructed position propagator:
\bea 
\label{e.dppgbc}
& \ds
P_{jk}^{\alpha \beta}(p^2) = 
{i \pi \over  v^2 (1 - q^2)} q^{-k-j}
\nl \ds
{[A^\alpha[0]  J_1({q^{-j} x} ;q^2) - B^\alpha[0] Y_1({q^{-j} x};q^2)]
[A^\beta[L]  J_1({q^{-k} x};q^2)  -  B^\beta[L]  Y_1({q^{-k} x};q^2)]
\over  A^\alpha[0] B^\beta[L] - B^\alpha[0] A^\beta[L]} \, .
\nn
\eea
Neumann (N) or Dirichlet (D) boundary conditions specify the parameters in \eref{dppgbc} as follows  
\bea & \ds
A^{\mathrm N}[0] =  Y_0( x;q^2) 
\qquad \qquad 
B^{\mathrm N}[0] = J_0(x;q^2) 
\nl \ds 
A^{\mathrm N}[L] =  Y_0(q^{-N-1} x;q^2) 
\qquad \qquad 
B^{\mathrm N}[L] = J_0(q^{-N-1} x;q^2)  
\nl \ds 
A^{\mathrm D}[0] =  Y_1(q x;q^2) 
\qquad \qquad 
B^{\mathrm D}[0] = J_1(q x;q^2) 
\nl \ds
A^{\mathrm D}[L] =  Y_1(q^{-N-1} x;q^2) 
\qquad \qquad 
B^{\mathrm D}[L] = J_1(q^{-N-1} x;q^2)  \, .
\eea
The correspondence between eqs. \erefn{ppgbc} and \erefn{dppgbc} can be established using the methods discussed in \sref{di}.
In particular, below the deviation scale $\Lambda_D = g v q^N$ 
(translated to the IR cutoff scale $\Lambda(L) = a_L \Lambda$),  physical amplitudes in deconstruction mimic those in the continuum theories at the leading order in $\delta$ and $p/gv$.  
Above the deviation scale, the spectra (the poles of the propagator) are different. 
However $t$-channel amplitudes on the UV brane are still reproduced in deconstruction, all the way to the UV cutoff  $\Lambda(0) = \Lambda$.  
Therefore for general boundary conditions the correspondence between deconstruction and continuum  holds at the same quantitative level as  for the Neumann--Neumann case discussed in Sections 3 and 4.

\section{Q-tutorial} 
\label{a.qb} 
\setcounter{equation}{0} 
 
This appendix contains definitions of  the q-Bessel functions and a review of their vital properties. 
Our  presentation is based mainly on ref. \cite{Sthesis}. 
We also derive several results concerning the asymptotic behaviour of the q-Bessel functions, which have not been given in the literature.   

Several inequivalent q-analogues of the Bessel functions have been
studied in the mathematical literature. The ones
relevant to our purpose are solutions
of the so-called Hahn--Exton q-difference equation, 
\beq
\label{e.hee}
(q^{\nu/2} + q^{-\nu/2} - q^{-\nu/2} t^2) F(t) 
- F(t q^{-1/2}) - F(t q^{1/2}) = 0 \, .
\eeq
For definiteness we consider $0<q<1$.  
One solution of this equation is the Hahn-Exton q-Bessel function~\cite{HE} (q-Bessel in the
following) denoted by $J_\nu(t;q)$.
It is defined by the power series
\beq
\label{e.qbd}
J_\nu(t;q) = t^\nu 
{(q^{\nu+1};q)_{\infty} \over (q;q)_\infty}
\sum_{k =0}^\infty {(-1)^k q^{k(k+1)/2} \over (q^{\nu+1};q)_{k}
  (q;q)_{k}}  t^{2k} \, ,
\eeq
where the q-shifted factorials are defined as  
\beq
(a;q)_k = \left\{ \begin{array}{ll} 1 & \mbox{if $k=0$} \\ 
 \prod_{n=0}^{k-1}(1 - a q^n)  & \mbox{if $k\geq 1$} \end{array}
\right. \, 
\eeq
for $a\in \mathbb{C}$ and $k \in \mathbb{Z}_+\equiv \{0,1,2,\ldots
\}$, and $(a;q)_\infty = \lim_{k\rightarrow \infty} (a;q)_k$. 
The q-Bessel $J_\nu(.;q)$ is analytic in $\mathbb{C}\backslash\{0\}$. 
For $\nu=n\in \mathbb{Z}$ and $|t|<1$, it has an integral representation:
\beq
J_n(t;q) = {1 \over 2 \pi} \int_0^{2 \pi} 
{(q t e^{-i \phi};q)_{\infty} \over (t e^{i \phi};q)_{\infty} } e^{- i
  n \phi} \, .
\eeq
It is known that all the zeros of the q-Bessel function of order
$\nu>-1$ are real and that the non-zero ones are simple~\cite{kost}. 


For $\nu \notin {\bf Z}$ $J_\nu(t;q)$ and $J_{-\nu} (t q^{-\nu/2};q)$ are two independent solutions of \eref{hee}; however, for integer $\nu = n$, there is the relation 
$J_{-n}(t;q) = (-1)^n q^\frac{n}{2} J_n (t q^\frac{n}{2};q)$.  
Therefore one introduces\footnote{%
Our definition differs by a factor $q^{-\frac{\nu^2}{2}}$ from
the one in~\cite{qNeumann}. With our definition,  $J$ and $Y$ satisfy
the same recurrence relations.}
 the q-Neumann function $Y_\nu(t;q)$, which is an independent solution for arbitrary $\nu$:
\beq
\label{e.qnd}
Y_\nu(t;q) =  {\Gamma_q(\nu) \Gamma_q(1- \nu)
  \over \pi} \, q^{-\frac{\nu^2}{2}} \, 
\left [ 
\cos (\pi \nu) q^{\nu \over 2} J_\nu(t;q) - J_{-\nu}(t q^{-
  \frac{\nu}{2}};q)  \right ] \, ,
\eeq 
where 
\beq
\Gamma_q(\nu) = {(q;q)_{\infty} \over (q^\nu;q)_\infty} (1 -
  q)^{1-\nu}  
\eeq
is a q-extension of the Euler $\Gamma$-function, satisfying 
$\lim_{q\to 1^-} \Gamma_q(\nu) = \Gamma(\nu)$.
For integer $\nu$, the limit $\nu \to n \in {\mathbb Z}$ is
  understood on the 
r.h.s. of \eref{qnd}. Explicitly, this limit gives~\cite{qNeumann}
\begin{align}
\label{e.nqn}
Y_n(t;q) = & \frac{2(q-1)}{\pi \log q}  J_n(t;q) \log
\frac{t}{1-q} \nn
& \mbox{} - \frac{1-q}{\pi}  t^{-n}
\sum_{k=0}^{n-1}\frac{(q;q)_{n-k-1} t^{2k}}{(q;q)_k}  \nn
& \mbox{} + \frac{1-q}{\pi \log q}  t^n
\sum_{k=0}^\infty 
    \frac{(-1)^k  q^{k(k+1)\over2} t^{2 k}}{(q;q)_k (q;q)_{n+k}} 
    \left\{\frac{\Gamma^\prime_q(n+k+1)}{\Gamma_q(n+k+1)}+
    \frac{\Gamma^\prime_q(k+1)}{\Gamma_q(k+1)} \right\} \nn
& \mbox{} -  \frac{1-q}{2\pi} t^n \sum_{k=0}^\infty
    \frac{(-1)^k  q^{k(k+1)/2} t^{2 k} (2k+1)}{(q;q)_k (q;q)_{n+k}} \,
\end{align}
for $n \in \mathbb{Z}_+$ (omitting the second term when $n=0$). 
For negative integers  
$Y_{-n}(t;q) = (-1)^n q^\frac{n}{2} Y_n (t q^\frac{n}{2};q)$. 

We now review properties of the q-Bessel and q-Neumann functions.  
Since the form relevant to deconstruction is $J_\nu(t;q^2)$ and
$Y_\nu(t;q^2)$ (rather than $J_\nu(t;q)$ and $Y_\nu(t;q)$), we find it
convenient to present all formulas in this form.  

The q-Bessels and q-Neumanns are q-analogues of the ordinary Bessel
and Neumann  functions,  
in the sense that there exists the  ``continuum limit'' $q \to 1$, 
\beq
\lim_{q\rightarrow 1^-} Z_\nu \left((1-q) t;q^2\right) =
Z_\nu(t) \, , \label{e.contJ}
\eeq
with $Z=J,Y$ and $J_\nu(.), Y_\nu(.)$ the ordinary Bessel and Neumann
functions.  
For $q$ close to $1$, the continuum approximation holds for $(1-q)
|t| \ll 1$, as can be seen from both the series and the 
integral representations. 
In fact, when $q = 1 - {\delta} \sim 1$ and $|t| \ll 1$, we can
Taylor expand $F(q^{\pm 1}t)$ in powers of $({\delta} t)$.  
Keeping terms up to the second derivative, we obtain  
\beq
{\delta}^2 t^2 F''(t) + {\delta}^2 t F'(t) + (q^{-\nu} t^2 -
q^{\nu} - q^{-\nu}) F(t) = 0 \, ,
\eeq
which is solved by the Bessel or Neumann functions:
\beq
\label{e.aaux1}
F(t) = Z_{4 \sinh^2 [\log q \nu/2]} \left ( {t \over q^\nu
  {\delta}} \right )  
\approx   Z_{\nu} \left ( {t \over {\delta}} \right )  \, .
\eeq 
This derivation is valid  as long as higher derivatives
in the Taylor expansion of  $F(q^{\pm 1}t)$ can be neglected, that is
  when $|{\delta} t F'''(t)| \ll |F''(t)|$. 
  For the solution \eref{aaux1} this is indeed true for $|t| \ll 1$.

The q-Bessels and q-Neumann satisfy recursion relations analogous to
the ones of their continuous cousins. In particular, the following
ones will prove very useful
\begin{align}&
\label{e.qbrr} 
q^\nu Z_\nu(t;q^2) - Z_\nu(q t;q^2) =  -  q t  Z_{\nu+1}(q t;q^2) \, ,
\nl
Z_\nu(t;q^2) - q^\nu Z_\nu(q t;q^2) =    t  Z_{\nu - 1}(t;q^2) \, ,
\end{align}
with $Z=J,Y$. These are actually discrete versions of
difference--recurrence relations. Indeed, one can define the
q-derivative of a function as
\beq
D_q f(t) = \left\{ \begin{array}{ll}\frac{f(t)-f(t q)}{t(1-q)} &
  ~\mbox{if}~t\neq 0  \\ f'(0) & ~\mbox{if}~t=0 \end{array} \right. \,
,
\eeq
which is a q-analogue of the continuum derivative: $\lim_{q\rightarrow
  1^-} D_q f(t) = f'(t)$. Then, \refeq{e.qbrr} are a
particular case ($k=1$) of~\cite{Sthesis}:
\begin{align} 
\label{e.diffrec} 
& \left[t^{-1} D_{q} \right]^k \left[t^{-\nu} Z_\nu(t;q^2)\right] =
\frac{(-1)^k t^{-\nu-k} q^{k(1-\nu)}}{\left(1-q \right)^k}
  Z_{\nu+k}(t q^k;q^2)   \nn
& \left[t^{-1} D_{q} \right]^k \left[t^\nu Z_\nu(t;q^2)\right] =
\frac{t^{\nu-k}}{\left(1-q\right)^k} Z_{\nu-k}(t;q^2) \, . 
\end{align}
For $q \to 1$ this reduces to the familiar recursion relations of the
continuum Bessel functions.  

The Wronskian of a differential equation has also its q-analogue. 
In general for a q-difference equation of the form  $F(q t) + F(q^{-1} t) + P(t,q) F(t) = 0$, its two independent $F_1(t)$ and $F_2(t)$ solutions satisfy the relation
$F_1(qt) F_2(t) - F_1(t) F_2(q t) = C(q)$   with the r.h.s. independent of $t$. 
For q-Bessels, the q-Wronskian relation is
\beq
\label{e.qbmwr} 
J_\nu(q t; q^2)Y_\nu( t; q^2) - Y_\nu(q t; q^2)J_\nu(t;q^2)
= {q^{-\nu} (1 - q^2) \over \pi} \, ,
\eeq
which can be proved using the series expansion of the q-Bessels.  
The fact that the q-Wronskian does not vanish at any point $t$
implies 
that a general solution of the q-difference equation~\refeq{e.hee} can
be written as~\cite{qNeumann}
\beq
F(t) = A(t) J_\nu(t;q^2) + B(t) Y_\nu(t;q^2) \, ,
\label{e.generalsolJY} 
\eeq
where
$A(t)$ and $B(t)$ are q-periodic, i.e. they satisfy $A(qt)=A(t)$, $B(qt)=B(t)$. 
But since we are interested only in
discrete values of the argument, $t_0, qt_0, q^2 t_0, \ldots$, we can
simply treat $A$ and $B$ as constants.

We can also define q-analogues of the modified Bessel and
Bessel--Macdonald functions, which are solutions of the Euclidean
version of the q-difference equation~\refeq{e.hee},
\beq
\label{e.heeEucl}
(q^{\nu/2} + q^{-\nu/2} + q^{-\nu/2} t^2) F(t) 
- F(t q^{-1/2}) - F(t q^{1/2}) = 0 \, .
\eeq
We define the q-modified-Bessel as\footnote{Here
  and in the following $i$ represents
  $e^{\pi i/2}$ when $-\pi < \mathrm{arg}\,t \leq \pi/2$.}
\beq
\label{e.mqbfk}
I_\nu(t;q) = i^{-\nu} J_\nu(i t;q) \, ,
\eeq
and the q-Bessel--Macdonald as
\beq
\label{e.mqbsk}
K_\nu(t;q) = \oh q^{-{\nu^2 \over 2}} \Gamma_q(\nu) \Gamma_q(1-\nu)
\left[ I_{-\nu}(q^{-{\nu \over 2}} t;q) -  q^{\nu \over 2} I_\nu(t;q)
  \right] \, ,
\eeq
with a limit understood for integer $\nu$. Our definitions agree
with the ones in~\cite{qEuclidean}, up to a change of variables and a
global factor. Equivalently, we can write
\beq
K_\nu(t;q) = \frac{\pi}{2} i^{\nu+1} \left[ \frac{\Gamma_q(\nu)
    \Gamma_q(1-\nu)}{\pi} \sin \nu \pi q^{\oh \nu(1-\nu)} J_\nu(it;q)
  + i Y_\nu(it;q)  \right] \, .
\eeq
Taking the limit of integer index, we find
\beq
K_n(t;q) =  \frac{\pi}{2} i^{n+1} \left[
\frac{(-1)^n  (1-q) (q;q)_{n-1}}{\log q (q^{1-n};q)_{n-1}}
q^{\oh n(1-n)} J_n(it;q) + i Y_n(it;q) \right] \, ,
\eeq
for $n\in \mathbb{N}$ and
\beq
K_0(t;q) =  \frac{\pi}{2} i \left[
\frac{(1-q)}{-\log q}
J_0(it;q) + i Y_0(it;q) \right] \, .
\eeq
The continuum limit is
\beq
\lim_{q\rightarrow 1^-} X_\nu \left((1-q) t;q^2\right) =
X_\nu(t) \, , \label{e.contK}
\eeq
with $X=I,K$.
The q-modified-Bessels and the q-Bessel--Macdonalds satisfy the
recurrence relations
\begin{align} 
\label{e.diffrecIK} 
& \left[t^{-1} D_{q} \right]^k \left[t^{-\nu} X_\nu(t;q^2)\right] =
\frac{(-1)^{\alpha k} t^{-\nu-k} q^{k(1-\nu)}}{\left(1-q \right)^k}
  X_{\nu+k}(t q^k;q^2)   \nn
& \left[t^{-1} D_{q} \right]^k \left[t^\nu X_\nu(t;q^2)\right] =
  (-1)^{\alpha k} \frac{t^{\nu-k}}{\left(1-q\right)^k}
  X_{\nu-k}(t;q^2) \, ,
\end{align}
with $\alpha=0,1$ for $X=I,K$, respectively.
The q-Wronskian is
\beq
\label{e.qbmwrEucl} 
I_\nu(q t; q^2) K_\nu( t; q^2) - K_\nu(q t; q^2) I_\nu(t;q^2)
= -{q^{-\nu} (1-q^2) \over 2} \, ,
\eeq
and a general solution of \refeq{e.heeEucl} can be written as
\beq
F(t) = A(t) I_\nu(t;q^2) + B(t) K_\nu(t;q^2) \, ,
\label{e.generalsolIK} 
\eeq
with $A,B$ q-periodic functions, which again we can consider
constant. 

The behaviour of all these functions at small values of the argument
can be read from the corresponding series expansions. 
We list  below the small-argument asymptotics that are useful for our
computations 
\begin{align}
\label{e.smallarg}
& J_0(t;q^2) \approx I_0(t;q^2) \approx 1 \, , \nn
& J_1(t;q^2) \approx I_1(t;q^2) \approx \frac{t}{1-q^2} \, , \nn
& Y_0(t;q^2) \approx \frac{1-q^2}{-\pi \log q} \left(\log
\frac{q^{\oh} t}{1-q^2} + \gamma_{q^2}\right) \, , \nn
& Y_1(t;q^2) \approx - \frac{1-q^2}{\pi} t^{-1} \, , \nn
& K_0(t;q^2) \approx \frac{1-q^2}{2 \log q} \left( \log \frac{q^{\oh}
  t}{1-q^2} + \gamma_{q^2} \right) \, , \nn
& K_1(t;q^2) \approx \frac{1-q^2}{2} t^{-1} \, , \nn
\end{align}
for $|t| \ll 1$. We have defined the q-Euler--Mascheroni constant
$\gamma_q=-\Gamma^\prime_q(1)$. Its series representation is
\beq
\label{e.qeulermascheroni}
\gamma_{q^2} = \log (1-q^2) - \log q^{2} \sum_{k=1}^\infty
\frac{q^{2k}}{1-q^{2k}} \, .
\eeq
Varying $q$ from 0 to 1, $\gamma_{q^2}$ grows approximately linearly
from 0 to $\gamma=0.577\ldots$.


Obtaining the asymptotic behaviour at large argument is more
involved. We first look
at the asymptotic form of the Hahn--Exton 
equation~\refeq{e.hee}. Writing $F(t) = \Pi(t) G(t)$, where 
\beq
\Pi(t) = \prod_{k=1}^\infty (1 - q^{2k}t^2) \equiv (q^2
t^2;q^2)_{\infty} \, 
\eeq
and using $\Pi(q t) = \Pi (t)/(1- q^2 t^2)$,  $\Pi(q^{-1} t) = \Pi
(t)(1- t^2)$ the Hahn--Exton equation takes the form 
\beq
G(q t) + (1 - t^2)(1- q^2 t^2) G(q^{-1} t) + (1 - q^2 t^2) (t^2
q^{-\nu} - q^{\nu} - q^{-\nu}) G(t) = 0 \, ,
\eeq 
which for $|t| \gg 1$ reduces to  
\beq
\label{e.aaux2}
G(q t) +  q^2 t^4  G(q^{-1} t)  -  q^2 t^4 q^{-\nu}  G(t) = 0 \, .
\eeq
One solution follows from neglecting the first term: 
$G(t) = {\rm const} \  t^\nu $.
From \eref{aaux2} we see that this approximation should work fine very
quickly when $|t|$ becomes larger than $1$. To find the second solution
it is more convenient to insert $F(t) = (1/\Pi(t)) G(t)$ in \eref{hee}, which
leads to the equation 
\beq
G(q t)(1 - t^2)(1- q^2 t^2) + G(q^{-1} t) + (1 -  t^2) (t^2 q^{-\nu} -
q^{\nu} - q^{-\nu}) G(t) = 0 \, ,
\eeq  
approximated by  $G(q t) =   q^{-\nu - 2} G(t) = 0$ at large $|t|$. The
solution is $G(t) =  {\rm const} \  t^{-\nu -2}$.   
Therefore, a general solution to the Hahn--Exton equation has the
asymptotic form 
\beq
F(t) \approx A t^\nu (q^2 t^2;q^2)_\infty
+ B {t^{-\nu -2} \over (q^2 t^2;q^2)_\infty} \, , \quad
\qquad |t| \gg 1 \, .   \label{e.genas}
\eeq
For imaginary $t$ one solution grows rapidly (faster than any power of
$t$) while the other  
decays (equally rapidly). 
For real $t$ the situation is more complicated as the ``decaying
part'' has poles at $t_n = q^{-n}$. 
In the following we pinpoint the coefficient $A$ for the q-Bessel functions defined previously.

We can obtain the asymptotic behaviour of the q-Bessel by approximating
$(q^{\nu+1};q)_k \approx (q^{\nu+1};q)_\infty$ in the power
 series~\refeq{e.qbd}. This is justified because for $t\gg 1$ the series is
 dominated by terms with large $k$ for which this
 approximation is fine.
  Using then the
so-called q-binomial theorem (see for instance the second reference
in~\cite{HE}), 
\beq
(-t^2;q^2)_\infty = \sum_{n=0}^\infty \frac{q^{n(n-1)}
  t^{2n}}{(q^2;q^2)_n}  \, ,
\eeq
we arrive at 
\beq
\label{e.qbltl}
J_\nu(t;q^2) \approx t^\nu \frac{(q^2
  t^2;q^2)_\infty}{(q^2;q^2)_\infty} \, ,~~|t| \gg 1 \, .
\eeq
This fixes the normalization of the growing part of the
q-Bessel. 
From this expression we can directly read the asymptotic zeros of the
q-Bessel $J_\nu(t;q^2)$, which are located at 
\beq
\label{e.qbltlz}
t_r \approx  q^{-r} \, ,~~ r\in \mathbb{N} \, , ~~q^{-r} \gg 1 \, ,
\eeq
independently of the index $\nu$ (though, in fact, the
approximation~\refeq{e.qbltl} is 
better for larger values of $\nu$). 
Since the above derivation is valid for any argument of $t$, we also
find 
\beq
\label{e.qIltl}
I_\nu(t;q^2) \approx t^\nu \frac{(-q^2
  t^2;q^2)_\infty}{(q^2;q^2)_\infty} \, ,~~ |t| \gg 1 \, .
\eeq 
Finding the asymptotic form of the q-Neumann is even more involved.
For integer index $n\in \mathbb{Z}_+$, using~\refeq{e.qnd}
  and~\refeq{e.qbltl} we have
\begin{align}
Y_n(t;q^2) & \approx  \frac{(-1)^n  (1-q^2) (q^2;q^2)_{n-1}}
{2 \pi \log q (q^2;q^2)_\infty (q^{2-2n};q^2)_{n-1}} 
\left[ q^{n(1-n)} \log(q t) t^n
  (q^2 t^2;q^2)_\infty   \phantom{\sum_{r=1}^\infty} \right. \nn  
  & \left.  \mbox{} - (-1)^n \left(\log \left (\frac{q^{2n}}{t} \right
  )+ 2 \log q 
 \sum_{r=1}^\infty \frac{t^2 q^{2(r-n)}}{1-q^{2(r-n)} t^2} \right)
 t^{-n}(q^{2-2n}t^2;q^2)_\infty \right] \, ,~|t| \gg 1 \, ,
\end{align}
for $n\in \mathbb{N}$,
\beq
Y_0(t;q^2) \approx \frac{(1-q^2)}{\pi \log q^{-1} (q^2;q^2)_\infty }
 \left[\log t+ \log q \left(\oh -
 \sum_{r=1}^\infty \frac{t^2 q^{2r}}{1-q^{2r} t^2} \right)\right]
 (q^{2}t^2;q^2)_\infty  \, ,~~ |t| \gg 1 \, .
\eeq
Note that the poles in the sum cancel against zeros in
$(q^{2-2n}t^2;q^2)_\infty$. For real $t$, writing $t=q^{-s+\alpha}$,
with $s\in \mathbb{N}$ and $\alpha \in ]-1/2,1/2]$, and assuming
$|\alpha|\ll 1$, we can approximate the sum at large $t$ by
\beq 
\sum_{r=1}^\infty \frac{t^2 q^{2(r-n)}}{1-q^{2(r-n)} t^2} =
\frac{\log t}{\log q} +1 - n + \frac{1}{q^{-2\alpha}-1} +
\mathcal{O}\left(t^{-2}\right) \, .
\eeq
Therefore, 
\begin{align}
Y_n(t;q^2) & \approx  \frac{(-1)^n  (1-q^2) (q^2;q^2)_{n-1}}
{2 \pi \log q (q^2;q^2)_\infty (q^{2-2n};q^2)_{n-1}} 
\left[ q^{n(1-n)} \log(q t) t^n
  (q^2 t^2;q^2)_\infty   \phantom{\frac{1}{q^{-2\alpha}-1}}
  \right. \nn   
  & \left.  \mbox{} - (-1)^n \left(\log t + 2
  \left(1+\frac{1}{q^{-2\alpha}-1}\right) \log q  \right)
 t^{-n}(q^{2-2n}t^2;q^2)_\infty \right] , \, |\alpha| \ll 1 , \, t \gg
1 \, ,  
\end{align}
for $n\in \mathbb{N}$, and
\beq
\label{e.qYltl}
Y_0(t;q^2) \approx \frac{(1-q^2)}{\pi (q^2;q^2)_\infty }
 \left(\frac{1}{2} + \frac{1}{q^{-2\alpha}-1} \right)
 (q^{2}t^2;q^2)_\infty  \, ,~~|\alpha|\ll 1 \, ,~ t \gg 1 \, .
\eeq

The corresponding expressions for
imaginary argument $t=i \tau$, $\tau \in \mathbb{R}$, are simpler. In
this case we can 
evaluate the relevant sum to leading order at large $|\tau|$ without
any further restriction: 
\beq 
-\sum_{r=1}^\infty \frac{\tau^2 q^{2(r-n)}}{1+q^{2(r-n)} \tau^2} =
\frac{\log \tau^2}{2\log q} - n + \frac{1}{2} +
\mathcal{O}\left(\tau^{-2}\right) \, ,
\eeq
which leads to the asymptotics
\beq
Y_n(i t;q^2) \approx i \frac{(-1)^n  (1-q^2) (q^2;q^2)_{n-1}
  q^{n(1-n)}} 
  {2\log q (q^2;q^2)_\infty (q^{2-2n};q^2)_{n-1}}  (i t)^n (-q^2
t^2;q^2)_\infty \, ,~~ t \in \mathbb{R} \, , ~~ |t| \gg 1 \, , 
\eeq
for $n\in \mathbb{N}$ and
\beq
Y_0(i t;q^2) \approx i \frac{(1-q^2)}{2\log q^{-1}(q^2;q^2)_\infty}  (-q^2 t^2;q^2)_\infty
\, ,~~ t \in \mathbb{R} \, , ~~ |t| \gg 1 \, . 
\eeq
This allows us to build a linear combination of the q-Bessel and
q-Neumann, which is purely decaying at large imaginary values of the
argument. In fact, from the above expression we see that this
combination is proportional to the q-Bessel--Macdonald. In other words, 
\beq
\label{e.qKltl}
K_n(t;q) \ll 1 \, ,~~ t \in \mathbb{R} \, , ~~ t \gg 1 \, . 
\eeq
Therefore, the q-Bessel--Macdonald shares with its continuum
counterpart the property that, at large real argument $t$, it is
damped. This is true  in the ``continuum'' region as well as for $t \gg 1$ 
where the  continuum approximation fails. We also see that in this
regime 
$K_n(t;q)\ll I_n(t;q)$. 
It is plausible that the same behaviour holds for non-integer indices,
though we have not proved this result.

\end{document}